\documentclass[useAMS,usenatbib]{mnras}
\pdfoutput=1
\usepackage[T1]{fontenc}
\usepackage{amsmath}
\usepackage{amssymb}
\usepackage{graphicx}
\usepackage{wasysym}
\usepackage{esint}
\usepackage[authoryear]{natbib}
\usepackage[utf8]{inputenc}
\usepackage{threeparttable}
\usepackage{hyperref}
\usepackage{xspace}
\usepackage{url}
\usepackage{babel}
\usepackage{ulem}
\usepackage{newtxtext,newtxmath}
\usepackage{hyperref}	
\usepackage{mathtools}
\usepackage{xcolor}
\hypersetup{colorlinks=true,linkcolor=blue,citecolor=blue,filecolor=blue,urlcolor=blue,}

\newcommand{\arepo}{\textsc{Arepo}\xspace}

\newcommand{\dd}{\mathrm{d}}
\newcommand{\bnabla}{\boldsymbol{\nabla}}
\newcommand{\msun}{\mathrm{M}_\odot}
\newcommand{\bs}[1]{\boldsymbol{#1}}
\newcommand{\bcdot}{\boldsymbol{\cdot}}

\newcommand{\DD}{\boldsymbol{\mathsf{D}}}
\newcommand{\crayon}{\textsc{Crayon+}\xspace}
\newcommand{\crest}{\textsc{CREST}\xspace}
\DeclarePairedDelimiter\abs{\lvert}{\rvert}

\makeatletter

\pdfpageheight\paperheight
\pdfpagewidth\paperwidth

\title[Gamma-ray emission from spectrally resolved CRs]{Gamma-ray emission from spectrally resolved cosmic rays in galaxies}

\author[M. Werhahn et al.]{Maria Werhahn,$^{1,2}$\thanks{E-mail:
mwerhahn@mpa-garching.mpg.de} Philipp Girichidis$^{3}$, Christoph Pfrommer,$^{1}$ Joseph Whittingham$^{1,4}$
\\
\\
$^{1}$Leibniz-Institut f\"ur Astrophysik Potsdam (AIP), An der Sternwarte 16, 14482 Potsdam, Germany\\
$^2$Max-Planck-Institut f\"ur Astrophysik (MPA), Karl-Schwarzschild-Str. 1, 85748 Garching, Germany\\
$^3$Institute for Theoretical Astrophysics (ITA), Albert-Ueberle-Str. 2, 69120 Heidelberg, Germany\\
$^4$Institut f\"ur Physik und Astronomie, Universit\"at Potsdam, Karl-Liebknecht-Str.\,24/25, 14476 Golm, Germany\\
}

\makeatother

\begin{document}
\date{Accepted 20XX . Received 20XX}

\maketitle

\label{firstpage}
\begin{abstract}
  Cosmic rays (CRs) are ubiquitous in the interstellar medium (ISM) of nearby galaxies, but many of their properties are not well-constrained. Gamma-ray observations provide a powerful tool in this respect, allowing us to constrain both the interaction of CR protons with the ISM and their transport properties. To help better understand the link between observational signatures and CR physics, we use a series of magneto-hydrodynamical (MHD) \arepo simulations of isolated galaxies performed using spectrally-resolved CR transport in every computational cell, with subsequent gamma-ray emission calculated using the \crayon (Cosmic RAY emissiON) code. In each of our simulated halos, modelling the energy-dependent spatial diffusion of CRs leads to a more extended distribution of high-energy (\textasciitilde100~GeV) gamma rays compared to that predicted by a `grey' steady-state model, which is especially visible in the corresponding emission maps and radial profiles.
  Despite this, the total gamma-ray spectra can often be well approximated by the steady-state model, although recovering the same spectral index typically requires a minor variation of the energy dependence of the diffusion coefficient. Our simulations reproduce the observed spectral indices and gamma-ray spectra of nearby star-forming galaxies and also match recent observations of the far infrared--gamma-ray relation. We find, however, that the spectrally resolved model yields marginally smaller luminosities for lower star formation rates compared to grey simulations of CRs. Our work highlights the importance of modelling spectrally resolved CR transport for an accurate prediction of spatially resolved high-energy gamma-ray emission, as will be probed by the upcoming Cherenkov Telescope Array observatory.
\end{abstract}

\begin{keywords}
  diffusion -- MHD -- methods: numerical -- cosmic rays -- galaxies: formation -- gamma-rays: galaxies
\end{keywords}

\section{Introduction}

The unexpected low efficiency with which galaxies form stars is still one of the biggest puzzles in galaxy formation \citep[e.g.\ ][]{1998Fukugita, 2010Moster}. It requires the existence of so-called feedback mechanisms that are able to quench star-formation by, e.g., expelling gas from the disc via galactic winds \citep[e.g.,][]{2017NaabOstriker}. While active galactic nuclei (AGNs) have been suggested to be an influential feedback mechanism in very massive galaxies and galaxy clusters \citep{2006Croton}, feedback from stellar winds, radiation fields and/or supernovae might be relevant in galaxies with masses of the order of the Milky Way and below. One of the potential mediators of this feedback are cosmic rays (CRs), a non-thermal relativistic population of charged particles, which have been found to be in rough equipartition with the thermal, magnetic and turbulent energy densities in the interstellar medium (ISM) of the Milky Way \citep[][]{1990BoularesCox}.

The main acceleration site of CRs are shocks that form at supernovae remnants (SNRs), whose occurrence is naturally closely connected to the star formation activity of a galaxy. We therefore expect a close connection between the star formation rate (SFR) and the CR content of a galaxy. This can be indirectly constrained through the observation of non-thermal emission processes from CRs, which arise across a wide range of frequencies. The radio band traces the underlying CR electron population through synchrotron radiation and tightly correlates with the SFR \citep{1971VanDerKruit,1973VanDerKruit,2003Bell,2021Molnar}, however, drawing inferences about the CR population using this band is often complicated as radio emission depends on properties of the galactic magnetic field. This, in turn, grows via a galactic dynamo \citep[e.g.,][]{2023BrandenburgNtormousi} making numerical modelling a necessity \citep{2021WerhahnIII,2021Whittingham,2022Pfrommer}, and its observed properties often have large uncertainties surrounding them. By contrast, the gamma-ray regime circumvents this issue as it mainly probes the interaction of CR protons with the ambient gas; such interactions generate neutral pions, which decay almost immediately into gamma-ray photons. If CR protons efficiently lost all of their energy through this channel, however, galaxies would be CR proton calorimeters \citep{1994Pohl}. Such efficient loss would, in turn, imply that CRs cannot escape galaxies, thereby preventing them from playing a significant role in driving galactic winds.
In order to constrain the relevance of CR feedback in galaxies, we therefore require an accurate modelling of the CR proton population in the formation and evolution of galaxies in combination with a modelling of the resulting non-thermal emission. 

There are a number of observations that provide constraints on this modelling. The gamma-ray emission of nearby star-forming galaxies has been observed by, e.g., the \citet{2009VERITAS_M82}, the \citet{2018HESS_NGC253} and the \citet{2022FermiLAT}. Through these, a strong correlation between the gamma-ray luminosity and SFR -- or equivalently the far-infrared (FIR) luminosity, which is a good tracer of star-formation -- has been found \citep{2012AckermannGamma, 2016RojasBravo,2020Ajello, 2020Kornecki}, as was predicted in earlier work \citep{2007Thompson}. This suggests that the calorimetric model holds for starburst galaxies, while other losses of CR protons lead to a deviation from this relation for lower SFRs \citep{2007Thompson,2011Lacki,2014Martin,2017bPfrommer,2020Kornecki,2021WerhahnII}. In particular, CR transport processes like advection and diffusion might be responsible for the comparatively reduced gamma-ray emission. Another benchmark for theoretical models are observed gamma-ray spectra from individual galaxies, which provide a more detailed constraint on the underlying distribution of CR protons, as shaped by the interplay of cooling and escape losses under the assumption that the emission is dominated by hadronic processes. Gamma-ray emission from individual galaxies was predicted by a number of works \citep[e.g.\ ][]{2004Torres,2005Domingo-SantamariaTorres, 2008Persic, 2009deCeaDelPozo, 2010Rephaeli} and later confirmed by subsequent observations and models \citep[e.g.\ ][]{2011Lacki,2013Yoast-Hull, 2019Yoast-Hull, 2019Peretti, 2022Ambrosone}.

Most modelling to date has adapted one-zone models, where free parameters are fit to the observed data. Indeed, only recently has gamma-ray emission from CRs been explored using full magneto-hydrodynamics (MHD) simulations of isolated galaxies \citep{2017bPfrommer, 2019Chan, 2020Buck, 2021WerhahnII, 2022Nunez-Castineyra}. In these simulations, CRs are coupled to the equations of MHD and treated as a relativistic fluid with an effective transport coefficient. Cooling rates are, furthermore, chosen assuming a universal steady-state spectrum, and the system is evolved using only the CR energy density \citep{2017aPfrommer}. We refer to this approach throughout the rest of the paper as the grey approach. 

In contrast, in \citet{2020Girichidis,2022Girichidis} a novel implementation was introduced for the moving-mesh code \arepo in which the spectral CR energy distribution is represented by a full spectrum in each computational cell. This treatment enables a dynamical coupling of CRs with the gas while evolving the full CR spectrum in time, as well as including cooling and spectrally resolved transport of CRs by means of energy-dependent spatial diffusion.
To identify the observational signatures that arise from this method, we calculate in this work the gamma-ray emission due to neutral pion decay using spectrally resolved CR MHD simulations of isolated galaxies combined with the emission pipeline of the \crayon code \citep[see ][for the calculation of the gamma-ray emission]{2021WerhahnII}. We then compare this to a steady-state model of CR spectra, also obtained using the \crayon code \citep[see][ for the calculation of the steady-state CR spectra]{2021WerhahnI}, applying this method to both the spectrally resolved simulations and the grey approximation. Finally, we verify our results against several observations in the gamma-ray regime.

Our work is structured in the following way. We first explain our numerical methods and the simulation setup in Section~\ref{sec: numerical methods and simulations}. We then analyse the morphological features that arise from the spectrally resolved CR simulations and compare them with the steady-state and grey modelling approaches in Section~\ref{sec: morphological differences}. In Section~\ref{sec: spectral differences}, we analyse the differences observable in the CR proton and gamma-ray spectra in detail. 
In Section~\ref{sec: comparison to observations and interpretation of delta}, we compare our results with recent data for nearby star-forming galaxies, including their gamma-ray spectra and the FIR-gamma-ray relation, and analyse the impact of altering the energy dependence of the diffusion coefficient. Finally, in Sections~\ref{sec: discussion and caveats} and \ref{sec: discussion and conclusion}, we discuss our results and conclude.

\section{Numerical methods and simulations}\label{sec: numerical methods and simulations}
\subsection{Spectrally resolved CR treatment}
The simulations underlying this work are introduced in a companion paper \citep{2023Girichidis}. Here we only briefly describe the model details. The simulations were performed with the second-order accurate, moving mesh code \arepo \citep{2010Springel, 2016aPakmor, 2020Weinberger} and follow the evolution of magnetic fields using the ideal MHD approximation.  We employ the method of cell-centred magnetic fields in \textsc{Arepo} \citep{2011Pakmor} with an HLLD Riemann solver \citep{2005JCoPh.208..315M} to compute fluxes and the Powell 8-wave scheme \citep{1999JCoPh.154..284P} for divergence cleaning \citep{2013Pakmor}. We also include the one-moment CR hydrodynamics solver \citep{2017aPfrommer}, which was extended to include spectrally resolved CR hydrodynamics developed by \citet{2020Girichidis} and linked to \arepo in \citet{2022Girichidis}. We briefly review the numerical method and physical processes in the following but refer to \citet{2020Girichidis,2022Girichidis} for a more detailed description.

The spectral code solves the Fokker-Plank equation for CRs
\begin{align}
    \frac{\partial f^{(3\mathrm{D})}}{\partial t} &= - \bs{\varv} \bcdot \bnabla f^{(3\mathrm{D})} 
    + \bnabla \bcdot [\DD \bcdot \bnabla f^{(3\mathrm{D})}] 
    + \frac{1}{3} (\bnabla \bcdot \bs{\varv}) \frac{\partial f^{(3\mathrm{D})}}{\partial p}
    \nonumber\\
    &+ \frac{1}{p^2} \frac{\partial}{\partial p} \left( p^2 b_l f^{(3\mathrm{D})}  \right) + j,
\label{eq: Fokker-Planck equation}
\end{align}
which describes the evolution of the isotropic part of the distribution function $f^{(3\mathrm{D})}=f^{(3\mathrm{D})}(\mathbfit{x},\mathbfit{p},t) =\dd N/(\dd p^3\dd x^3)$ as a function of time, $t$, and $p=\abs{\mathbfit{p}}=P/(m_\rmn{p}c)$, the absolute value of the particle momentum normalised to the proton mass, $m_\rmn{p}$, times the speed of light, $c$. We neglect diffusion in momentum space and assume anisotropic spatial CR diffusion along the magnetic field using the discretisation of  \citet{2016cPakmor}. For this, we use the spatial diffusion tensor $\DD=D\bs{b}\bs{b}$, where $D=D(p)$ is the parallel diffusion coefficient with momentum dependence
\begin{align}
    D(p) = 10^{28} p^\delta \mathrm{cm^2\,s^{-1}},
    \label{eq:diff}
\end{align}
for which we choose $\delta=0.3$, and $\bs{b}=\bs{B}/B$ denotes the direction of the magnetic vector field, $\bs{B}$, with field strength $B=\sqrt{\bs{B}^2}$. For the remaining terms in Eq.~\ref{eq: Fokker-Planck equation}, $\bs{\varv}$ is the mean velocity of the thermal gas, and $j$ and $b_l=\dd p/\dd t$ are source and loss terms, respectively. For the latter, losses due to Coulomb and hadronic interactions of CR protons with the ambient medium are taken into account. In addition, we account for streaming losses using the simplified approach of \citet{2017aPfrommer} and \citet{2020Buck}, in which streaming drains energy from the CRs at a rate proportional to the Alfv\'{e}n speed $\bs{\varv}_\mathrm{A}$ and the CR pressure gradient $\bnabla P_\mathrm{cr}$. This amounts to a corresponding CR loss rate of $\Lambda_\mathrm{cr} = |\bs{\varv}_\mathrm{A}\bcdot\bnabla P_\mathrm{cr}|$. We apply the Alfv\'{e}n cooling to the total spectrally integrated CR energy in every cell individually. We then apply the change in CR energy by keeping the spectral shape and simply rescale the amplitude of the spectrum to yield the new total CR energy.

The particle distribution function is discretised in momentum space and is represented by a piece-wise power law. This provides two degrees of freedom: the normalisation and the slope of the spectrum in each momentum bin. The first two moments of the distribution function, i.e.\ the CR number and energy density, are chosen to be evolved in time for each momentum bin.

\subsection{Simulation setup}

We simulate the formation of galaxies with the moving-mesh code \arepo using the same setup as described in \citet{2022Girichidis}, but vary some of the parameters. We start with a dark matter halo that is characterised by an NFW profile and a concentration parameter, $c_{200}=7$, where we vary the halo mass $M_{200}=\{10^{10},10^{11},3\times 10^{11},10^{12}\}\,\msun$. The spin of the dark matter halo is parametrized by the spin parameter $\lambda = J \sqrt{\abs{E}} G^{-1} M_{200}^{-5/2}$, where $J$ denotes the total angular momentum, $G$ is the gravitational constant, and $E$ the total energy of the halo.
We chose $\lambda=0.3$ and adapt rigid body rotation with $\omega=j(r)/r^2 = \mathrm{const}$.
The dark matter halo contains gas that is initially in thermodynamic equilibrium but as soon as we start the simulation and switch on cooling, it collapses and forms a disc. Stars form in a stochastic manner following the \citet{2003SpringelHernquist} ISM model, where an effective equation of state is adopted based on the assumption that the hot and cold phase are in equilibrium.
We inject CRs with an efficiency of 10 per cent of the canonical SN energy of $10^{51}\,$erg.

We run all setups with two models of CR transport, respectively. In the first, we adopt the `grey' approach, where we follow only the evolution of the CR energy density using the advection-diffusion approximation introduced in \citet{2017aPfrommer}. In the second, we run a set of the same simulations but this time apply our novel spectrally resolved CR hydrodynamics solver, as summarised in the previous section \citep{2022Girichidis}. For this, we discretise the spectrum in twelve equally spaced logarithmic momentum bins ranging from $100\,\mathrm{MeV}\,c^{-1}$ to $100\,\mathrm{TeV}\,c^{-1}$. In the vicinity of newly formed star particles, we inject CRs with an injection spectrum $f^{(3\mathrm{D})}(p)\propto p^{-4.2}$. We summarise our set of simulations in Table~\ref{tab: Simulations}.

In order to better isolate the impact of employing a spectral treatment of CR proton transport and to better interpret the resulting gamma-ray emission, we follow three different approaches:
\begin{itemize}
    \item Model `grey': we apply the cell-based steady-state approach as introduced in \citet{2021WerhahnI} (see Section~\ref{sec: steady-state modelling}) to the grey runs.
    \item Model `spec': we directly take the CR spectra of the spectrally resolved CR runs and compute their resulting gamma-ray emission from neutral pion decay.
    \item Model `steady on spec': we apply the cell-based steady-state model (see Section~\ref{sec: steady-state modelling}) to the novel spectrally resolved CR runs, thereby obtaining steady-state CR proton spectra from the exact same simulations.
\end{itemize}
This last step is constructive because it enables a direct comparison between steady-state modelling and the new spectrally resolved simulations of CRs. This is because the new spectral treatment has a minor dynamical impact on the global evolution of the galaxy in comparison to the grey approach \citep{2023Girichidis}. Simply comparing the spectrally resolved CR simulations to the grey simulations could, therefore, dilute the interpretation of any potential differences.

\begin{table}
\renewcommand{\arraystretch}{1.5}
\begin{centering}
\caption{Overview of the simulations.}
\label{tab: Simulations}
\begin{tabular}{lcccc}
\hline
 $M_{200}\,[\mathrm{M_{\odot}}]$ & CR model & $D\,\mathrm{[cm^2 s^{-1}]}$ & name \\
\hline
$10^{10}$ &          grey         & $10^{28}$& \texttt{M1e10-grey} \\
$10^{11}$ &          grey         & $10^{28}$& \texttt{M1e11-grey} \\
$3\times 10^{11}$ &  grey         & $10^{28}$& \texttt{M3e11-grey} \\
$10^{12}$ &          grey         & $10^{28}$& \texttt{M1e12-grey} \\
$10^{10}$ &  spectrally resolved&
Eq.~\eqref{eq:diff}& \texttt{M1e10-spec} \\
$10^{11}$ & spectrally resolved& Eq.~\eqref{eq:diff}& \texttt{M1e11-spec} \\
$3\times10^{11}$ & spectrally resolved& Eq.~\eqref{eq:diff}& \texttt{M3e11-spec} \\
$10^{12}$ & spectrally resolved& Eq.~\eqref{eq:diff}& \texttt{M1e12-spec} \\
\hline
\end{tabular}
\par\end{centering}
\end{table}

\subsection{Steady-state modelling\label{sec: steady-state modelling}}
For the steady-state solution, we apply the \crayon post-processing code introduced in \cite{2021WerhahnI}. In this, we solve the diffusion-loss equation  \citep[see, e.g.,][]{1964ocr..book.....G, 2004Torres} for the one-dimensional distribution function 
\begin{equation}
f(E_\rmn{p})=f(p) \frac{\dd p}{\dd E_\rmn{p}} = 4\uppi p^2 f^{(3\mathrm{D})}(p) \frac{\dd p}{\dd E_\rmn{p}}
\end{equation}
for each computational cell. This means solving
\begin{align}
\frac{f(E_\rmn{p})}{\tau_{\mathrm{esc}}}-\frac{\mathrm{d}}{\mathrm{d}E_\rmn{p}}\left[f(E_\rmn{p})b(E_\rmn{p})\right]=q(E_\rmn{p}),
\label{eq: diffusion-loss-equation}
\end{align}
with source and loss terms $q$ and $b$, respectively. The escape losses due to diffusion and advection are quantified by an escape timescale $\tau_\rmn{esc}^{-1}=\tau_\rmn{diff}^{-1}+\tau_\rmn{adv}^{-1}$, where the latter timescales are estimated as
\begin{align}
    \tau_{\mathrm{diff}}=\frac{L_{\mathrm{CR}}^{2}}{D(E_\rmn{p})}\propto E_\rmn{p}^{-\delta},
    \qquad
    \tau_{\mathrm{adv}}=\frac{L_{\mathrm{CR}}}{\varv_{z}}, 
\end{align}
where $L_\mathrm{CR}=\varepsilon_{\mathrm{CR}}/\left|\bnabla\varepsilon_{\mathrm{CR}}\right|$ is the estimated diffusion length in each cell, and $\varv_z$ is the $z$-component of the gas cell velocity. This estimate assumes that advection is effectively only happening in the $z$-direction, implying that all fluxes in the azimuthal direction in and out of the cell compensate one another \citep[see figure~6 of][]{2021WerhahnI}. As a source term, we assume a power law in momentum $q(p)= q[p(E_\rmn{p})] \dd E_\rmn{p}/ \dd p $ with an exponential cut-off  
\begin{align}
q(p)\dd p = C_0 \, p^{-\alpha_{\mathrm{inj}}} \exp[- p/p_{\mathrm{cut}}] \dd p.
\label{eq: source function steady-state q(p)}
\end{align}
The cut-off momentum, $p_{\mathrm{cut}}$, is chosen to be $1\,\mathrm{PeV}/(m_\mathrm{p}c)$  \citep{1990Gaisser} if not mentioned otherwise, whilst the normalisation $C_0$ is determined such that the integral over the equilibrium distribution function $f$ matches the CR energy density $\varepsilon_\rmn{CR}$ of each cell, i.e. $\int E_\mathrm{kin}(p) f(p) \dd p = \varepsilon_\rmn{CR}$, where $E_\mathrm{kin}=(\sqrt{(p^2+1)} -1)m_\rmn{p} c^2$.

\subsection{Gamma-ray emission}
We calculate the gamma-ray emission arising from the hadronic interactions of CR protons with the ambient gas in our simulations using the emission pipeline of the \crayon code, as described in \citet{2021WerhahnII}. The source function for gamma-ray emission due to neutral pion decay, i.e. $q_E=\dd N_\gamma/(\dd V\,\dd t\,\dd E)$, is computed using the model by \citet{2018Yang} for energies ranging from the pion production threshold up to 10~GeV and by \citet{2014Kafexhiu} for larger proton kinetic energies. From this, we obtain the specific luminosity as an integral over the volume $V$, i.e.\ $L_E = \int q_E (\bs{r})\dd V$. The spectral surface brightness, $S_E$, and the spectral flux, $F_E$, are defined as
\begin{align}
  S_E(\bs{r}_\perp) = \int_{-\infty}^{\infty} q_E (\bs{r}) \dd l,
  \quad\mbox{and}
  \quad
  F_E = \frac{1}{4\uppi d^2} \int q_E (\bs{r})\dd V, 
\label{eq: S_E and F_E}
\end{align}
where $\bs{r}_\perp$ is the radius vector in the plane orthogonal to the projection direction and $d$ denotes the luminosity distance.
We furthermore define the integrated luminosity in the energy range from $E_1$ to $E_2$ as 
\begin{align}
L_{E_1-E_2}=\int_{\Omega} \dd V \int_{E_1}^{E_2} E q_E\, \dd E.
\label{eq: L_(E1-E2)}
\end{align}

\section{Morphological differences} \label{sec: morphological differences}

\begin{figure*}
\begin{centering}
\includegraphics[scale=0.97]{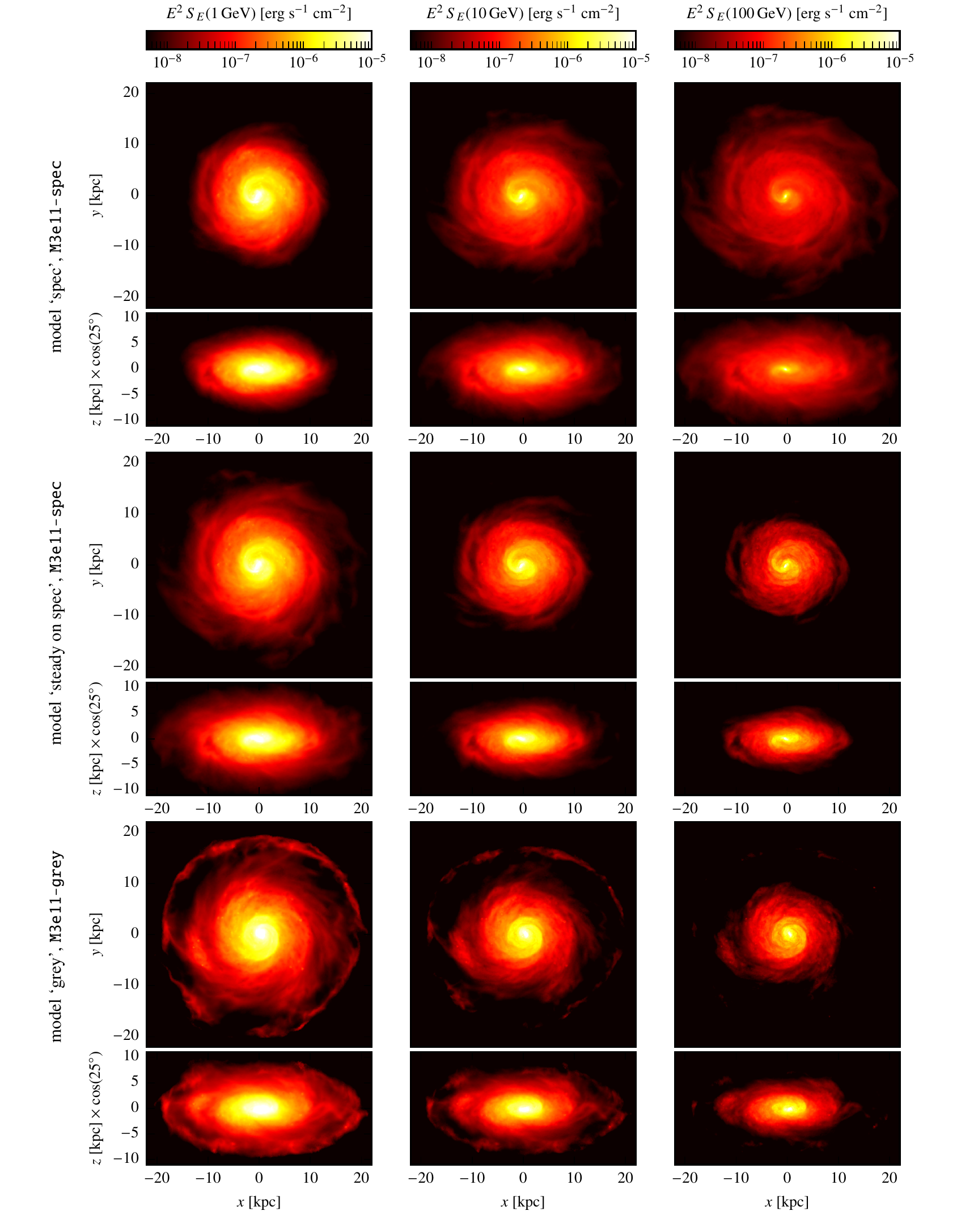}
\par\end{centering}
\caption{Gamma-ray emission from neutral pion decay of spectrally resolved CR protons (\texttt{M3e11-spec}), where energy dependent diffusion is included explicitly (upper six panels), and the emission resulting from steady-state CR spectra of the same simulation (middle six panels) at 1, 10 and 100~GeV (left-, middle and right-hand panels), respectively, for time $t=$1~Gyr. We show projected maps face-on (first, third and fifth row) as well edge-on views that are rotated by 25 degrees (second, forth and sixth row). The lower six panels show the same maps calculated from steady-state CRs of a simulation performed with the grey approach (\texttt{M3e11-grey}).} 
\label{fig:maps-gamma-ray-spec}
\end{figure*}

\begin{figure*}
\begin{centering}
\includegraphics[scale=1.]{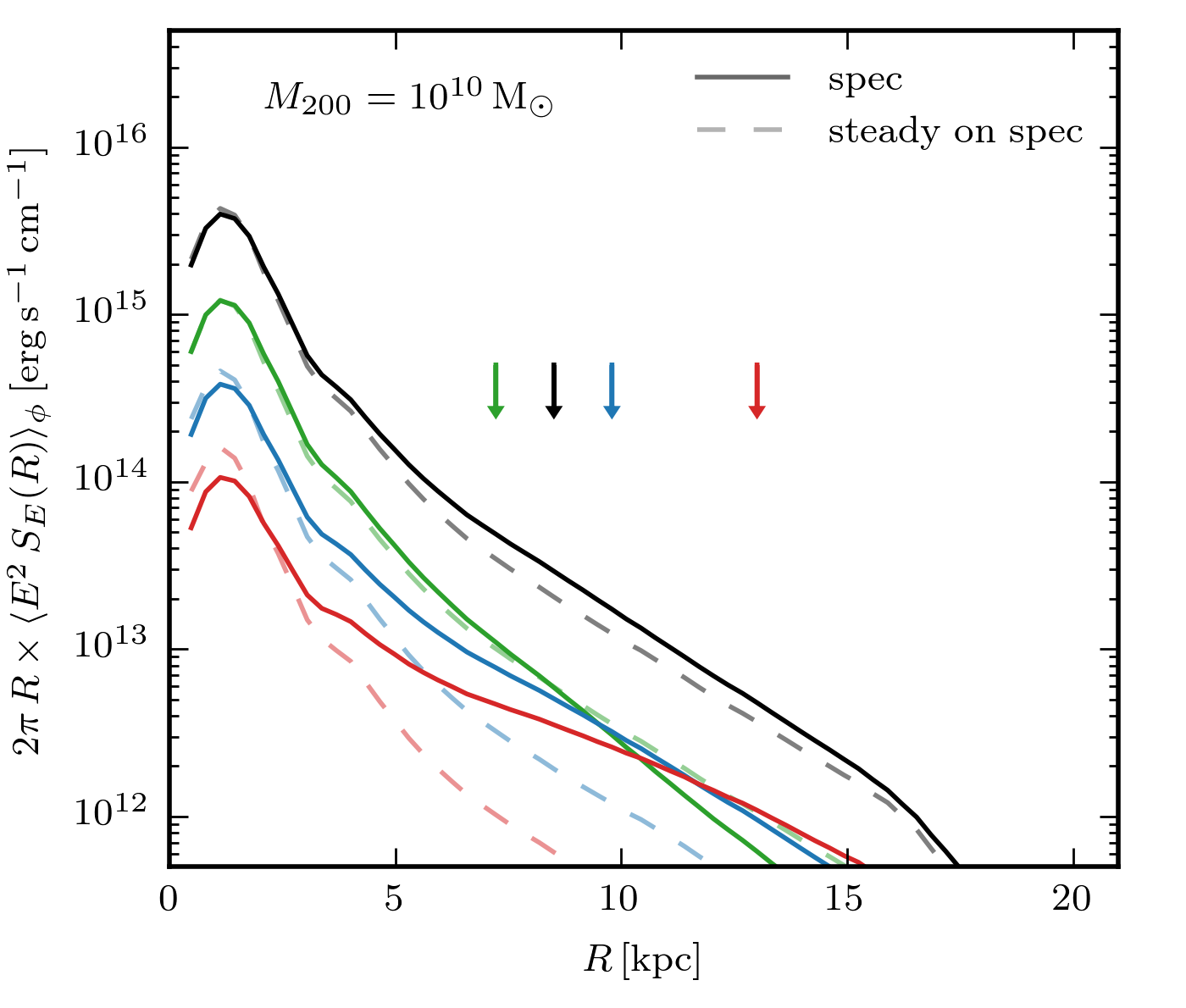}\includegraphics[scale=1.]{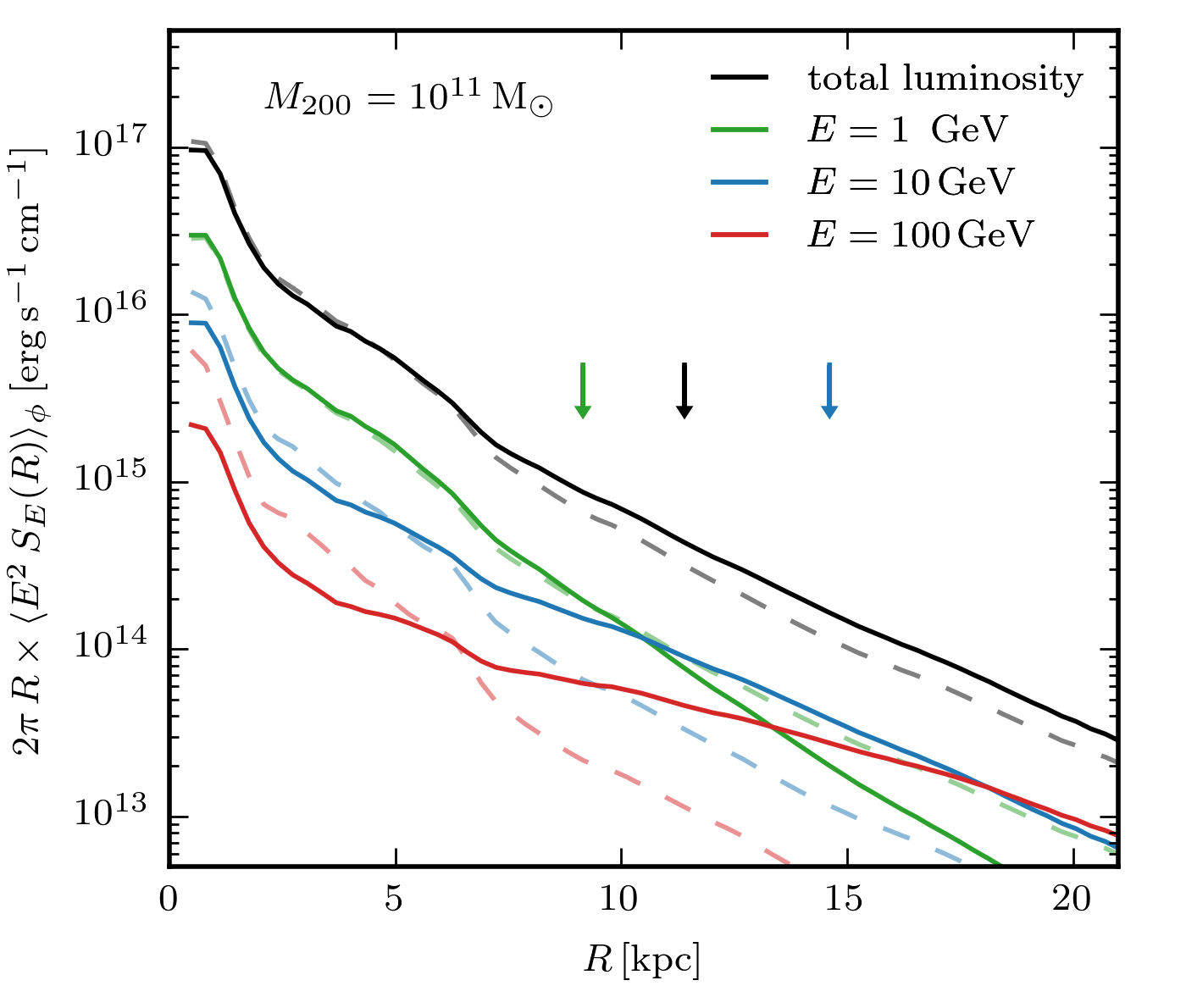}\\
\includegraphics[scale=1.]{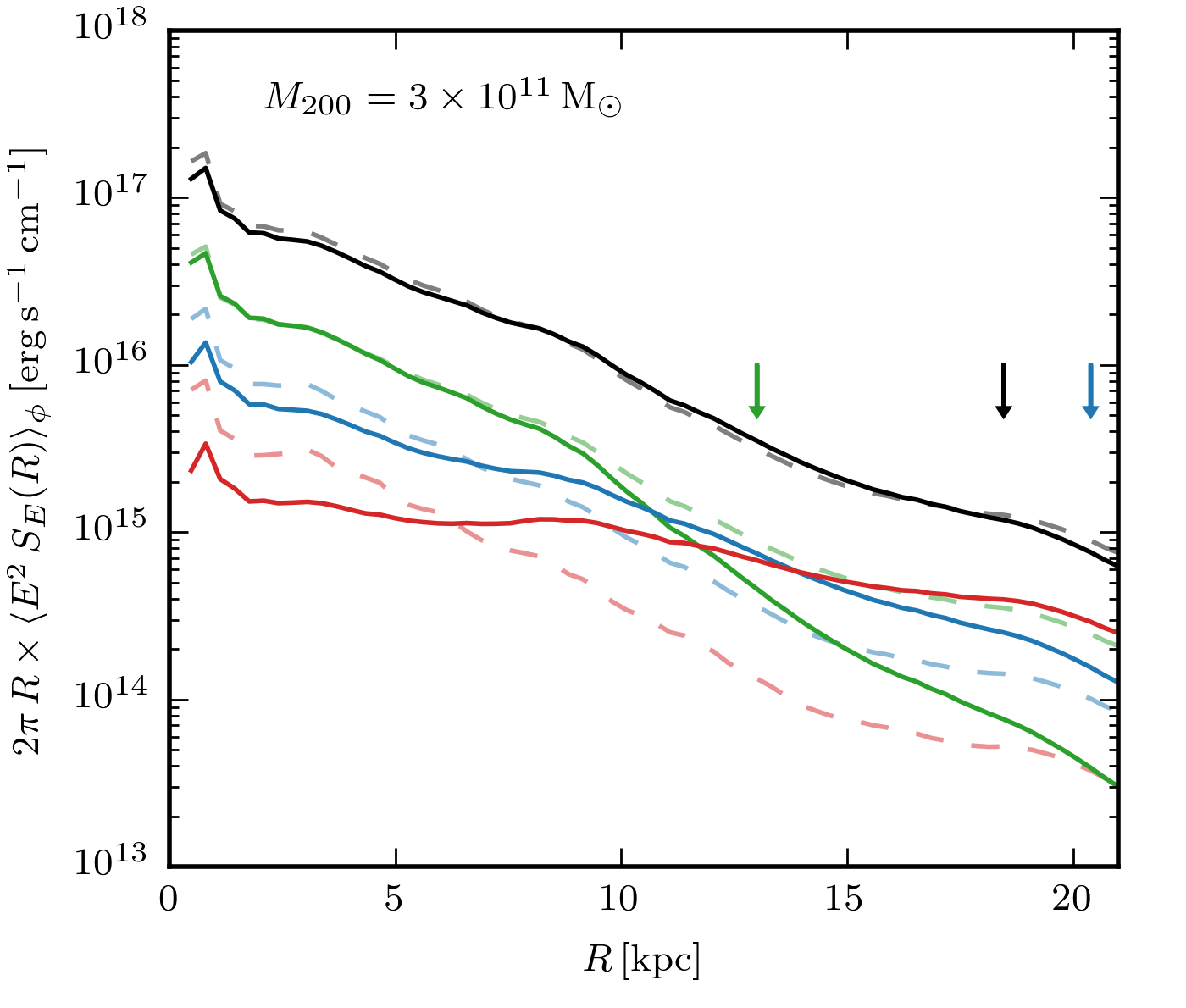}\includegraphics[scale=1.]{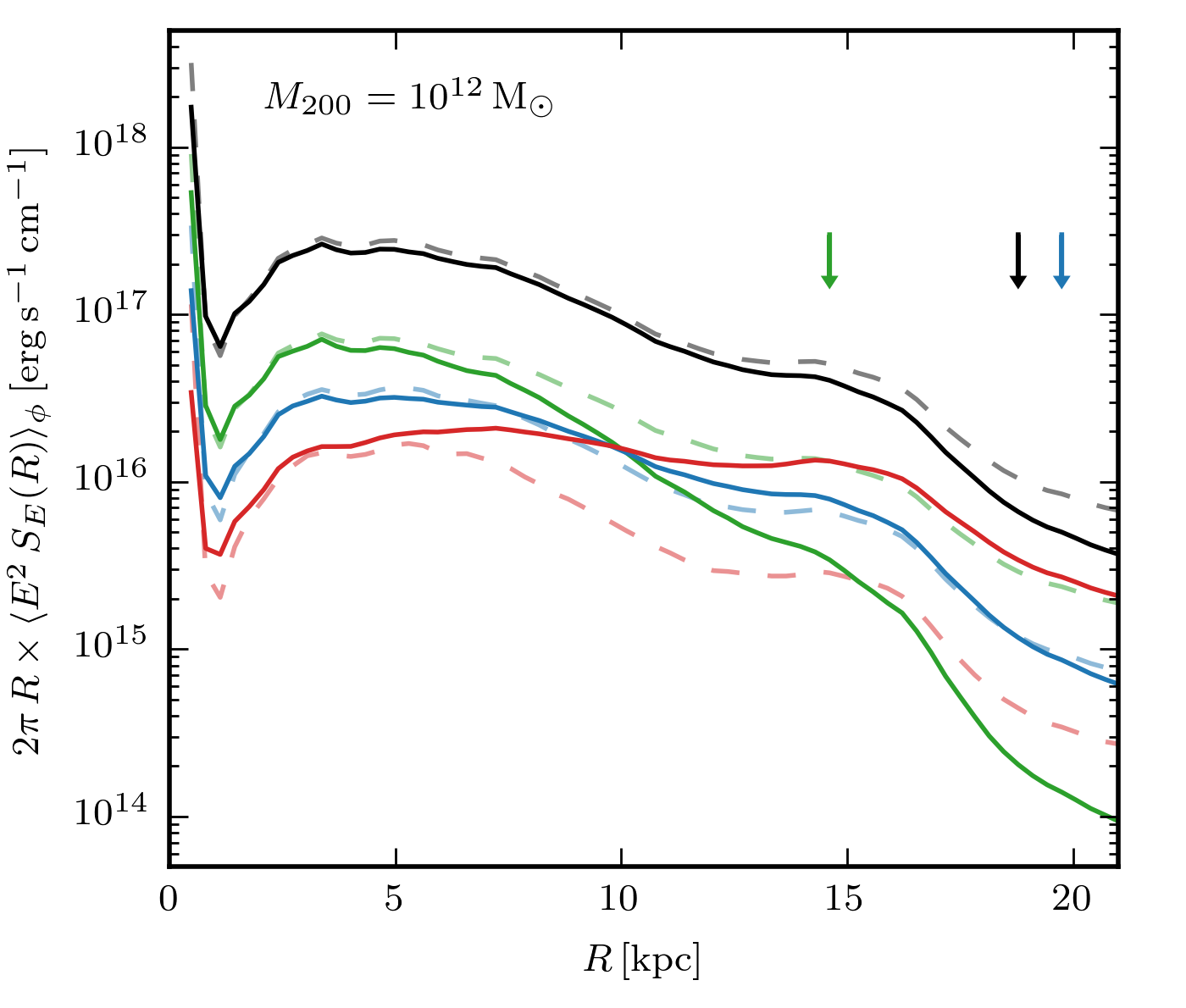}
\par\end{centering}
\caption{Radial profiles of the differential contribution to the gamma-ray emission for the different halo masses at $t=1$~Gyr for the spectrally resolved CR runs (solid lines) and the steady-state model applied to the same simulations (dashed lines). The different colours show $E^2 \mathrm{d}L_E/\mathrm{d} R = 2\uppi R\times \langle E^2 S_E(R)\rangle_\phi$ at different gamma-ray energies $E$ as indicated in the legend in the upper right-hand panel. The differential contribution for the integrated luminosity $L_{0.1-100\mathrm{GeV}}$ is shown in black, where 99 per cent of this luminosity is included within a radius that is represented by the black arrow for each halo, respectively. The radius where 99 per cent of the specific luminosity $L_E$ at a given energy is included is indicated by a vertical arrow in the corresponding colour for each $E$ in the `spec' model.} 
\label{fig:gamma-ray-diff-contribution}
\end{figure*}

The most striking effect of the spectrally resolved CR treatment on the gamma-ray emission from our simulated galaxies can be seen in Fig.~\ref{fig:maps-gamma-ray-spec}. Here, in the upper six panels, we present gamma-ray emission maps obtained directly from our spectrally resolved CRs simulation \texttt{M3e11-spec} (model `spec'), whilst in the middle six panels, we show the emission from the steady-state model as applied to the same simulation (model `steady on spec'). The projected maps are shown face-on and nearly edge-on for gamma-ray emission at 1, 10 and 100~GeV, respectively. It can be seen that in the spectrally resolved model the GeV-gamma-rays are more centrally concentrated compared to the emission from the steady-state model, whilst gamma-rays with energies of 10 or 100~GeV extend to comparatively larger radii. This happens because gamma-ray maps at 1 and 100~GeV probe CR protons that typically have energies of 100~MeV and 10~GeV, respectively. Following Eq.~\ref{eq:diff}, this equates to a difference in the diffusion coefficient by a factor of $100^{0.3}\approx 4$. Consequently, the inclusion of spectrally resolved CR transport allows high-energy CRs to diffuse faster than the steady-state model whilst low-energy CRs diffuse slower, allowing the 100~GeV emission to have a much greater radial extent in the face-on maps.
This effect can also be seen in the nearly edge-on maps, which reveal more extended emission at 10 and 100~GeV for the spectrally resolved CR model when compared to the corresponding steady-state approach.

The lower six panels of Fig.~\ref{fig:maps-gamma-ray-spec} show the simulation \texttt{M3e11-grey}, where the gamma-ray emission has been calculated from the steady-state model (model `grey'). 
There is no explicit energy-dependent diffusion in our grey simulations, and subsequently the high-energy CRs do not diffuse to the same extent as in the `spec' model.
This leads to a much more concentrated high-energy gamma-ray emission at 100~GeV in this model (cf. the right-hand panels in Fig.~\ref{fig:maps-gamma-ray-spec}). The steady-state approach yields this result for both the steady-state applied to the grey simulations and applied to the spectral runs. We only find slightly varying distributions in gas density. 

To quantify the morphological differences, we compute the differential contribution to the specific gamma-ray luminosity via
\begin{align}
E^2 \frac{\mathrm{d}L_E}{\mathrm{d}R}= E^2  \frac{d}{dR} \int_{\mathrm{d}\Omega} q_E R\, \dd R\, \dd \phi\, \dd z =  2\uppi R\times \langle E^2 S_E(R)\rangle_\phi ,
\end{align}
where the azimuthally averaged surface brightness is defined by
\begin{align}
\langle E^2 S_E(R)\rangle_\phi = \frac{E^2}{2\uppi}  \int_{-\infty}^{\infty} q_E  \dd \phi\,\dd z 
\label{eq: definition E S_E(R)}
\end{align}
and for the integrated gamma-ray luminosity 
\begin{align}
 \frac{\mathrm{d}L_{0.1-100\mathrm{GeV}}}{\mathrm{d}R} &=   \frac{d}{dR} \intop_{E_1}^{E_2} \intop_{\mathrm{d}\Omega} E q_E R\, \dd R\, \dd \phi\, \dd z \dd E\\
    &=  2\uppi R S_{0.1-100\mathrm{GeV}}(R),
\end{align}
where $R$ denotes the cylindrical radius, $E_1=0.1\,\mathrm{GeV}$ and $E_2=100\,\mathrm{GeV}$.

We show the differential contribution to the gamma-ray emission as a function of radius in Fig.~\ref{fig:gamma-ray-diff-contribution} for our spectrally resolved CR runs with different halo masses after 1~Gyr of evolution. We show the steady-state model applied to these simulations in dashed lines. It can be seen that, in this case, there is a very similar radial contribution to the gamma-ray emission from each energy $E$, for each radius and halo mass. This is not the case, however, for the spectrally resolved CR treatment (solid lines). Here, the 
contribution of the different gamma-ray energies to the overall emission varies with radius. This is due to the fact that the CR diffusion speed depends on energy.

The effect is in principle visible for all halo masses. However, while in our smaller galaxies \texttt{M1e10-spec} and \texttt{M1e11-spec} we find that 100~GeV gamma-rays only become dominant outside of the radius where 99 per cent of the total gamma-ray luminosity is included (black arrows), these gamma-rays already contribute substantially below that radius for larger halo masses. In particular, in our most massive halo with $M_{200}=10^{12}\,\msun$, 100~GeV gamma-rays dominate over all lower gamma-ray energies already at radii $\sim$10~kpc, while 99 per cent of the total, integrated gamma-ray luminosity is not reached until $R\sim$19~kpc.

In contrast to the profiles of the specific luminosities at different energies, the radial distribution of the total gamma-ray luminosity (black lines in Fig.~\ref{fig:gamma-ray-diff-contribution}) closely follow the profiles for a steady-state configuration in all considered halo masses. This is unsurprisingly as the (momentum-integrated) CR energy densities of both models are identical in each cell by construction and consequently, integrating the gamma-ray emission over an interval large enough to cover the majority of the pressure-carrying CR particle energies must reflect this.
We conclude, therefore, that while the radial distribution of the total gamma-ray luminosity is not sensitive to the underlying model of the CR spectra, the spatial gamma-ray distributions at different energy bands vary substantially and require resolving spectral CR transport in order to accurately capture their distribution.

\section{Spectral differences}\label{sec: spectral differences}

Having explored the spatial signatures of spectrally resolved CR transport, we now examine the spectral differences by analysing first the CR proton spectra, before inspecting the gamma-ray spectra.

\subsection{CR proton spectra}

\subsubsection{Steady-state and grey approach}

\begin{figure}
    \centering
    \includegraphics{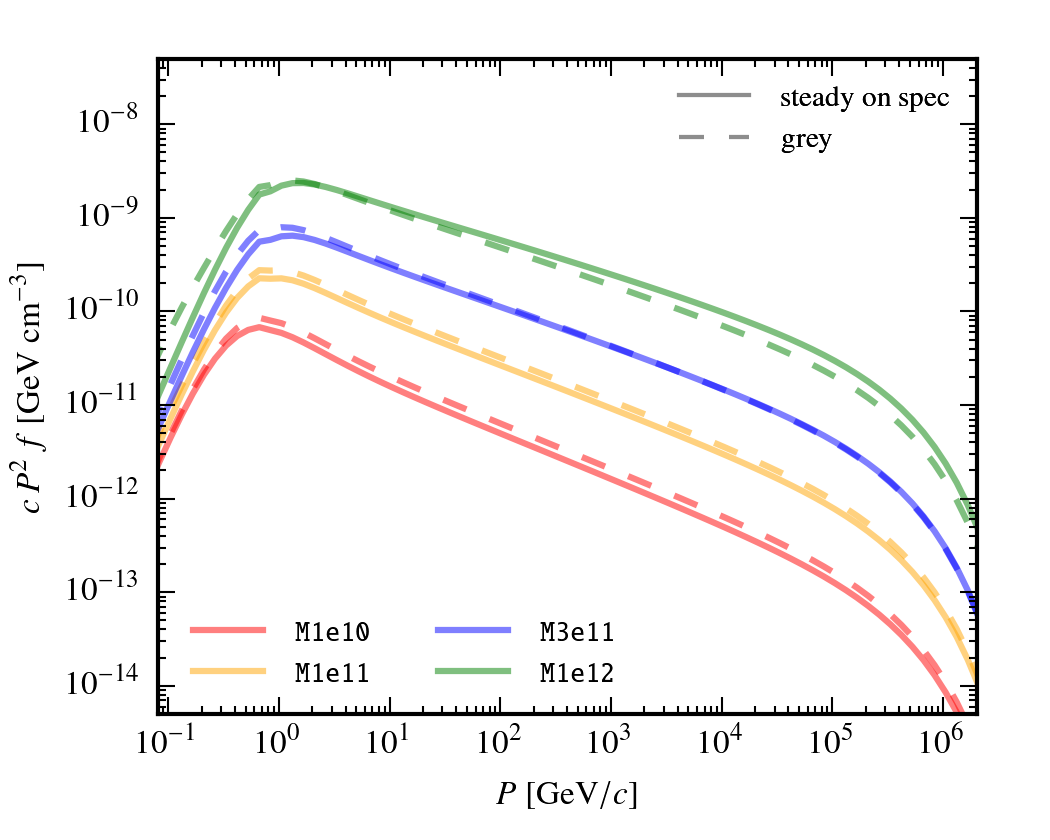}
    \caption{Steady-state CR proton spectra of all spectrally resolved (solid lines) and grey runs (dashed lines) after $t=1$~Gyr. For all halo masses (shown in different colors as indicated in the legend), the steady-state CR proton distributions (averaged over a cylinder with $R=20$~kpc and a height $h=1$~kpc above and below the midplane) are almost identical. }
    \label{fig: CR protons all halos}
\end{figure}

First, we compare the CR proton spectra of the same steady-state model applied to our grey and our spectral CR runs.
Figure~\ref{fig: CR protons all halos} shows the steady-state proton spectra of all our simulated halos after an evolution of 1~Gyr, averaged over a cylinder with $R=20$~kpc and a height $h=1$~kpc above and below the midplane. 
While we find some minor morphological differences in the gamma-ray emission arising in our two steady-state models (see middle and lowest panels of Fig.~\ref{fig:maps-gamma-ray-spec}) due to the moderately different dynamical impact of spectrally resolved CR hydrodynamics \citep{2023Girichidis}, Fig.~\ref{fig: CR protons all halos} shows that the averaged steady-state CR proton spectra are almost identical for both the grey and spectrally resolved runs for each halo mass. 
This means that we may reasonably compare the spectrally resolved CRs directly to the steady-state approach applied to the same runs, instead of comparing them to the grey simulations. Note, however, that employing the spectrally resolved model leads to differences in the morphology, the outflow efficiencies, and the CGM properties, which is analysed in the simulation paper \citep{2023Girichidis}, while we will focus here mainly on the properties of the emerging gamma-ray emission.

\subsubsection{Spectrally resolved CR protons vs. steady-state}\label{sec: spectral CRs vs. steady-state}

\begin{figure*}
\begin{centering}
\includegraphics[]{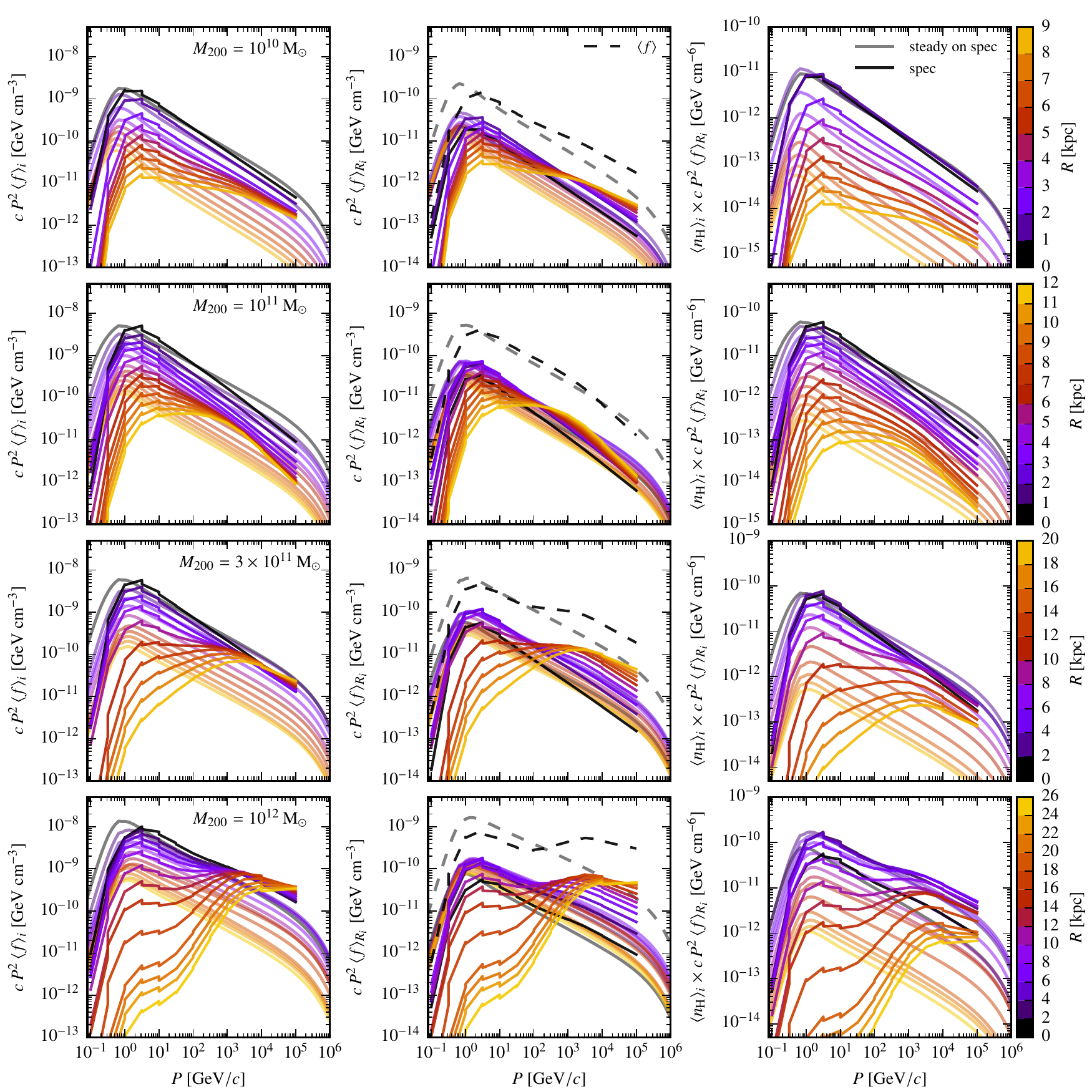}
\par\end{centering}
\caption{CR proton spectra for the different halo masses (increasing from top to bottom) at $t=1$~Gyr for the `spec' and `steady on spec' model, respectively. The coloured lines show the averaged CR spectra in concentric, cylindrical bins $i$ with a height $h= \pm 1\,\mathrm{kpc}$ from the mid plane, ranging from $R=0$ to $R=R_\mathrm{max}$ (indicated by different colours), where we define a maximum radius $R_\mathrm{max}$ such that the CR energy density has decreased approximately by four e-foldings at this radius. 
The left-hand panels show the CR spectra averaged in each radial bin $i$ (Eq.~\ref{eq: <f>_i}), while the middle panels show the contribution of each radial bin to the total spectrum (Eq.~\ref{eq: <f>_{R_i}}) where the spectrum averaged over the total volume is shown by the dashed lines for each model, respectively (Eq.~\ref{eq: <f> total}). In order to study the contribution of different radii to the total gamma-ray emission, we multiply the CR spectra in the right-hand panels by the averaged gas density $\langle n_\mathrm{H} \rangle_i$ of each radial bin.} 
\label{fig:CR-proton-spectra-radialbins}
\end{figure*}

The differences in the morphology of the gamma-ray emission found in Fig.~\ref{fig:maps-gamma-ray-spec} for the spectrally resolved CRs in comparison to the corresponding steady-state model can be traced back to differences in the underlying CR proton distribution. To enable a detailed comparison of the spectrally resolved CR approach with the steady-state approach, we show in the left-hand panels of Fig.~\ref{fig:CR-proton-spectra-radialbins} the CR proton spectra averaged in radial bins for our different halo masses. We obtain the averaged spectra from the spectrally resolved CR simulations by first averaging the CR energy and number density in each momentum bin over the region of interest and then reconstructing the normalisation and slope for each bin. The steady-state spectra are computed in post-processing for every computational cell for the same simulations (see Section~\ref{sec: steady-state modelling}) by assuming the same energy scaling of the diffusion coefficient ($\delta=0.3$) and an identical injected spectral index for the CR protons. We calculate the average spectra as
\begin{align}
    \langle f \rangle_i =  \frac{1}{V_i} \int f_i\, \dd V_i 
    \label{eq: <f>_i}
\end{align}
in each radial bin $i$ with volume $V_i$, respectively. 
There are only very few cases where we find the spectrally resolved treatment to yield a spectrum that resembles the corresponding steady-state spectrum in the left-hand panels of Fig.~\ref{fig:CR-proton-spectra-radialbins}. Only in the case of our dwarf galaxy with $M_{200}=10^{10}\,\msun$ in the radial region $1~\mathrm{kpc} \lesssim R \lesssim 3~\mathrm{kpc}$ do the CR spectra exhibit similar slopes comparable to the steady-state model albeit with a different cut-off at low momenta. This is due to the explicit modelling of cooling as the spectra evolve as a function of time, while the steady-state approach assumes injection and cooling to be in equilibrium at the time of the snapshot.
Apart from this case, the CR spectra in all other regions are not found to be close to a steady-state configuration.
Instead, they either have a similar shape with a steeper slope or a completely different distribution.
The former predominantly occurs in the central regions of all but the most massive simulated galaxy. After an initial injection of CRs in the central regions due to the high SFRs there, high energy CRs are allowed to diffuse out very quickly in the spectral code which leads to steep central spectra, while they diffuse into the outskirts where the CR injection rate is lower. This leads to flatter spectra at large radii compared to the steady-state model, where injection and cooling are always assumed to be in balance. This effect is particularly strong in the \texttt{M1e12-spec} simulation, where the spectra in the `spec' model exhibit positive slopes in the outskirts in comparison to the negative slopes of the steady-state spectra (in the `steady on spec' model) above $\sim1$~GeV.

However, the left-hand panels of Fig.~\ref{fig:CR-proton-spectra-radialbins} only represent a volume-weighted average of the CR proton spectra in each radial bin, respectively, which does not reflect the actual contribution to the total spectrum. Hence, we additionally show the contribution to the total spectrum in radial bins $R_i$ in the middle column of Fig.~\ref{fig:CR-proton-spectra-radialbins} where we calculate each contribution by
\begin{align}
\langle f\rangle_{R_i} = \frac{1}{V} \int f_i \dd V_i  = \frac{2 \uppi }{V} \intop_{R_{i-1/2}}^{R_{i+1/2}} f_i \ R_i\, \dd R\, \dd\phi\,\dd z.
\label{eq: <f>_{R_i}}
\end{align}
This definition ensures that the spectra fulfill $\Sigma_i \langle f\rangle_{R_i} = \langle f\rangle$ by construction, where the CR proton spectrum averaged over the total volume $V$ is given by
\begin{align}
\langle f\rangle = \frac{1}{V} \int f \dd V.  
\label{eq: <f> total}
\end{align}
Note that the definition of the average in Eq.~\eqref{eq: <f>_{R_i}} differs from that in Eq.~\eqref{eq: <f>_i} only by the normalising volume: in the latter case, the spectra are simple volume averages in the radial bins (left-hand panels of Fig.~\ref{fig:CR-proton-spectra-radialbins}) while Eq.~\eqref{eq: <f>_{R_i}} computes the actual contribution of each radial bin to the total spectrum averaged over the whole volume (middle panels of Fig.~\ref{fig:CR-proton-spectra-radialbins}).
However, with increasing radius the gas density steeply decreases and hence, this representation does not reflect the radial contribution of the CR proton spectra to the total gamma-ray spectrum because the source function of gamma-ray emission from neutral pion decay is directly proportional to the gas density $n_\mathrm{H}$.
Consequently, we additionally consider the quantity $\langle n_\mathrm{H}\rangle \times \langle f \rangle_{R_i}$ in order to be able to identify the relevant spectral features that impact the gamma-ray emission.
This is shown in the right-hand panels of Fig.~\ref{fig:CR-proton-spectra-radialbins} and will help us to interpret the differences in the gamma-ray emission in Section~\ref{sec: Gamma-ray spectra from spectrally resolved CRs}. 

Comparing the middle and right-hand panels of Fig.~\ref{fig:CR-proton-spectra-radialbins} reveals that the effect of high-energy CRs diffusing outwards more quickly (than expected in a steady-state) is visible in the total spectra of all halo masses (middle panels) but taking into account the gas density changes the picture (right-hand panels). In the latter, we clearly see that only the central few kpc will be relevant for the emerging gamma-ray emission, where the CR spectra of the \texttt{M1e10-spec} simulation closely resemble a steady-state, whereas the \texttt{M1e11-spec} simulation exhibits steeper CR spectra. In contrast, in the two most massive halos we expect an increasingly higher contribution from the relatively flat CR spectra due to energy-dependent diffusion. In particular, in the \texttt{M1e12-spec} simulation, this effect causes substantially harder gamma-ray spectra.

\subsection{Gamma-ray spectra from spectrally resolved CRs} \label{sec: Gamma-ray spectra from spectrally resolved CRs}

In this section, we aim to examine the effect of the spectrally resolved CR scheme on the resulting gamma-ray emission. 
First, we note that the steady-state modelling applied to the spectrally resolved simulations yields almost exactly the same spectral shapes in total gamma-ray emission as the steady-state applied to the grey runs for all halo masses at all times (see their corresponding CR proton spectra exemplified in Fig.~\ref{fig: CR protons all halos} at $t=1$~Gyr). 
There are only minor differences in the normalisation of the total gamma-ray spectra and their morphologies. These are principally due to corresponding variations in the star formation histories and gas distributions of the simulated discs, which arise due to the way spectrally-resolved CR transport affects the dynamics of the system \citep{2023Girichidis}.
Therefore, we again focus on the comparison of the emission from the spectrally resolved CRs to the steady-state approach applied to the same spectral runs in the following.

\subsubsection{Temporal evolution of gamma-ray spectra}

\begin{figure*}
\begin{centering}
\includegraphics[scale=1.]{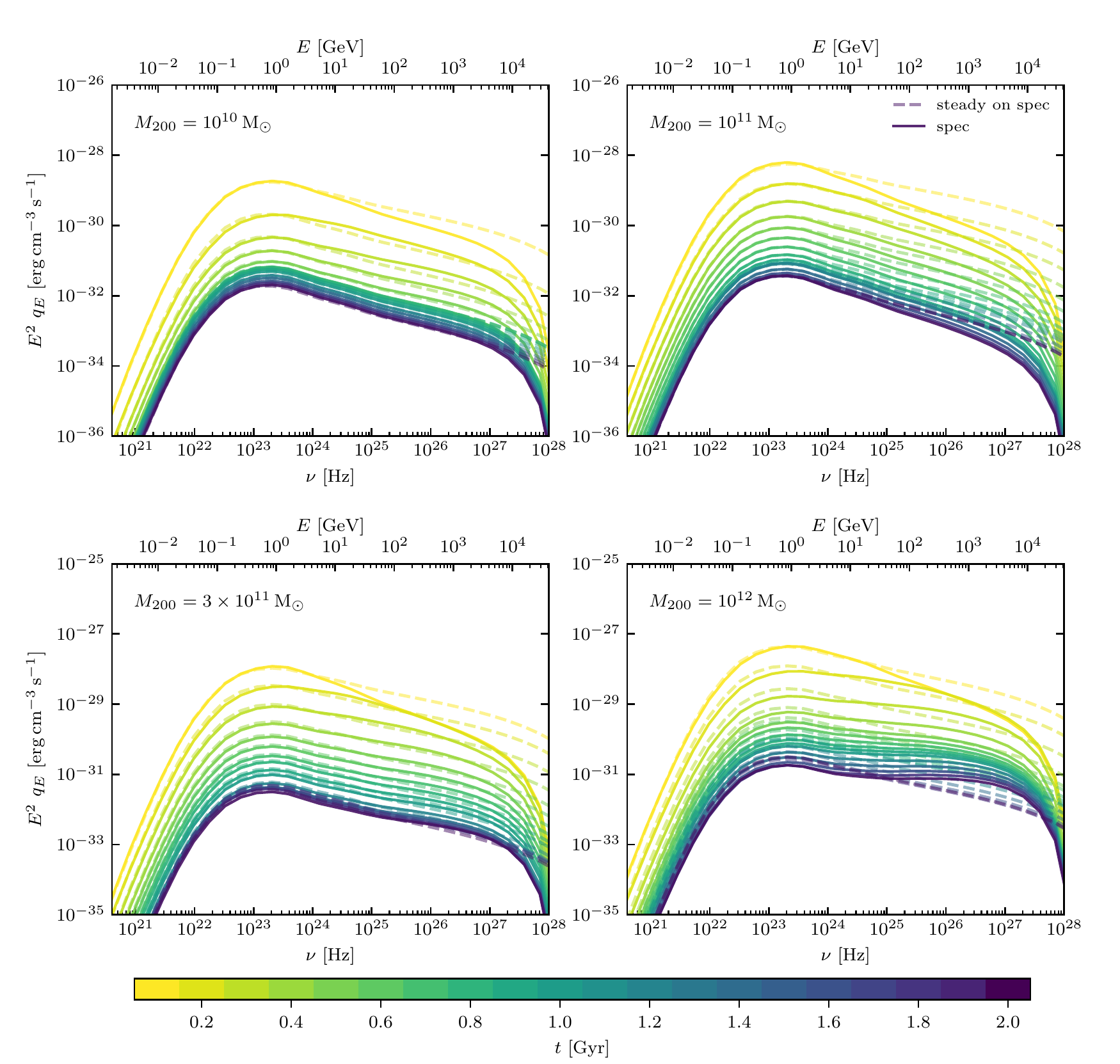}
\par\end{centering}
\caption{Time evolution of the total gamma-ray spectra from neutral pion decay for our simulated galaxies with different halo masses using the spectrally resolved CR method (model `spec', solid lines). The dashed lines show the resulting emission from the steady-state model applied to the same simulations (model `steady on spec'). Note that the spacing between the plotted spectra is $\Delta t=0.1$~Gyr up to 1~Gyr and $\Delta t=0.2$~Gyr afterwards for better visibility.}  
\label{fig:gamma-ray-spectra-time-evolution}
\end{figure*}

In Fig.~\ref{fig:gamma-ray-spectra-time-evolution} we show the temporal evolution of the total gamma-ray spectra from neutral pion decay for all our simulated galaxies in the spectrally resolved runs from $t=0.1$ to 2~Gyr for all simulated halos. 
The initial collapse of the gas cloud ignites a peak in star formation and hence results in a high injection of CRs. This decreases exponentially in time, which causes a corresponding decrease in the normalisation of the gamma-ray spectrum as a function of time in all simulations.

The comparison of the emission from the spectrally resolved CR protons with the steady-state model reveals that while the gamma-ray spectra match a steady-state configuration (i.e.\ our steady-state approach applied to the same simulations) in simulations \texttt{M1e10-spec} and \texttt{M3e11-spec} at late times, all halos exhibit steeper spectra at early times. Curiously, the spectra of the \texttt{M1e11-spec} simulation continue to be steeper even at later times, whilst the gamma-ray spectrum of the \texttt{M1e12-spec} halo quickly hardens with time and exceeds the corresponding steady-state spectra above $\sim 10$~GeV from $\sim 0.3$~Gyr onwards. To understand the reason for the deviations from a steady-state configuration outlined above, we examine the radial contributions to the total gamma-ray spectrum in the following.

\subsubsection{Radial contribution to gamma-ray spectra} \label{sec: Radial contribution to gamma-ray spectra}

\begin{figure*}
    \centering
    \includegraphics[]{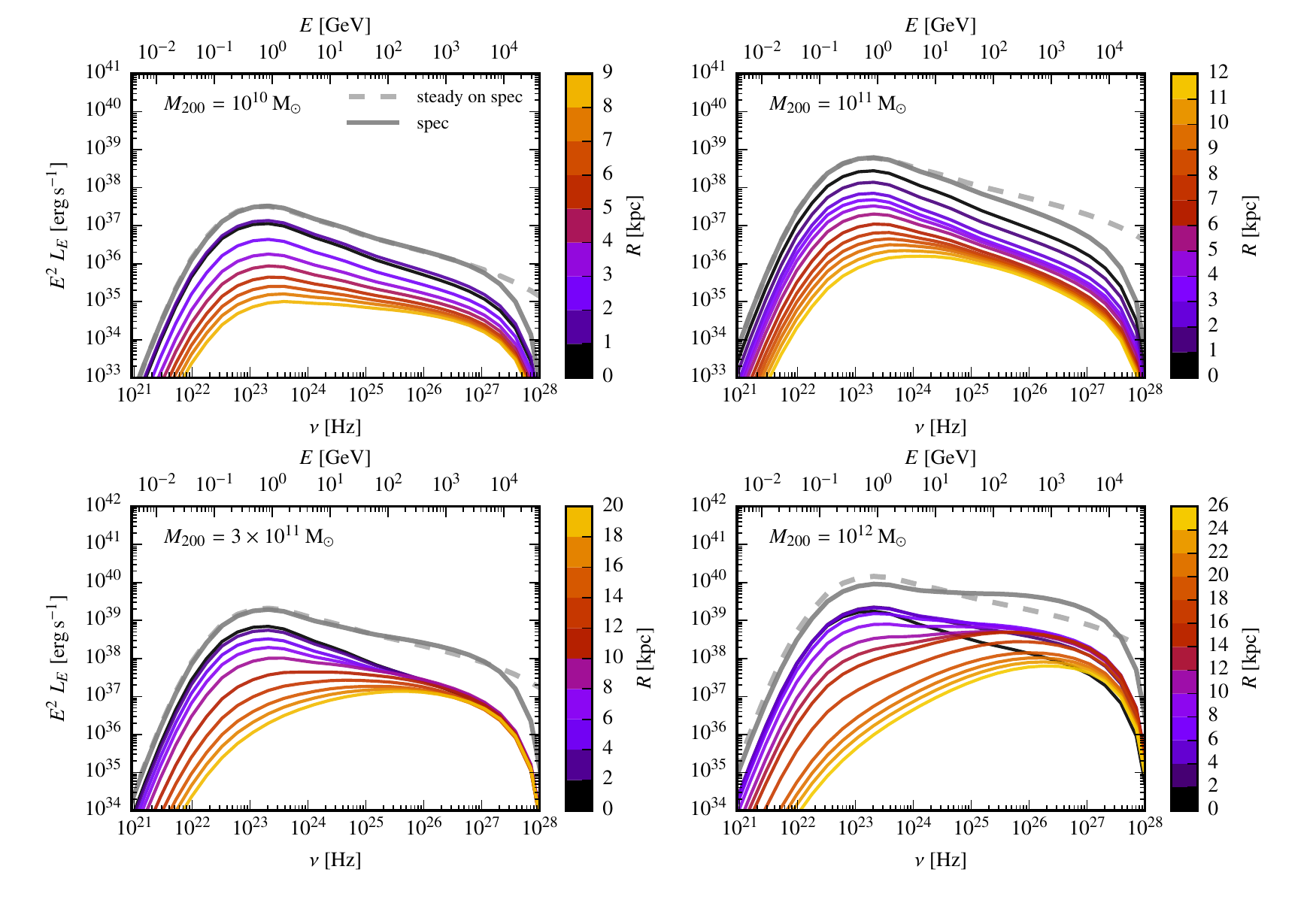}
    \caption{Gamma-ray spectra from neutral pion decay at $t=1$~Gyr for different halo masses. The total gamma-ray spectra from the steady-state CRs (model `steady on spec', dashed grey lines) are shown on top of the emission resulting from the spectrally resolved CR simulations (model `spec', solid grey lines). Additionally, we show the differential contribution to the latter from different radial, concentric bins, where we define a maximum radius such that the CR energy density has decreased by four e-foldings (i.e.\ the same radial bins as shown in Fig.~\ref{fig:CR-proton-spectra-radialbins}). }
    \label{fig: gamma-ray spectra radial bins}
\end{figure*}

The effect of explicit energy-dependent diffusion on the CR spectra in the spectrally resolved model (as discussed in Section~\ref{sec: spectral CRs vs. steady-state}) is imprinted in the radial contribution to the total gamma-ray spectra and hence varies strongly with halo mass. This is illustrated in Fig.~\ref{fig: gamma-ray spectra radial bins} where we show the gamma-ray spectra at $t=1$~Gyr for all halo masses in the `spec' model (solid lines) in radial bins together with the total spectrum obtained in the `steady on spec' model (dashed lines). The range of the concentric radial bins is chosen such that at the maximum radius, the CR energy density has decreased by approximately four e-foldings (i.e.\ we use the same radial bins as in Fig.~\ref{fig:CR-proton-spectra-radialbins}).

In the dwarf galaxy \texttt{M1e10-spec} we find the total gamma-ray spectrum to be dominated by the central region within $\sim 3$~kpc, where the spectral shape of the gamma-ray emission is very close to the spectrum from a steady-state configuration. This reflects the agreement we found in the corresponding CR proton spectra in Fig.~\ref{fig:CR-proton-spectra-radialbins} in the central region.
Only in the outskirts at larger radii are the gamma-ray spectra much harder. However, the contribution from those large radii is sub-dominant due to the decreasing gas densities with radius. This is evident from comparing the middle and right-hand panels of the upper row in Fig.~\ref{fig:CR-proton-spectra-radialbins} where we multiplied the averaged CR spectra by the gas density (as discussed in Section \ref{sec: morphological differences}).

However, the situation is very different in the simulation \texttt{M1e11-spec}. Here, the high-energy CRs diffuse outwardly more quickly than they can be replenished from fresh injection, which means that no steady-state configuration can be achieved and consequently, the total gamma-ray spectrum is steeper than in a steady-state.\footnote{Note that we adapt here for the steady-state modelling the same energy dependence of the diffusion coefficient ($\delta=0.3$) as in the spectrally resolved run. We explore the impact of varying $\delta$ in Section~\ref{sec: spectral index of gamma-ray emission}.} 
The outwards diffusing CRs lead to flatter CR proton spectra in the outskirts (see also Fig~\ref{fig:CR-proton-spectra-radialbins}). But due to the decreasing gas density in these regions, the gamma-ray emission is too low in normalisation in order to be able to compensate for the very steep central spectra that dominate the total emission.

This is in contrast to the next more massive halo \texttt{M3e11-spec}. Here, we first have the same situation that CRs diffuse outwards very quickly, which steepens the central CR proton and hence also the central gamma-ray spectrum (see Fig.~\ref{fig:CR-proton-spectra-radialbins} for the CR proton spectra). But here, the flat CR proton spectra in the outskirts arising from the outwards diffused CRs are in a region where the gas density is still high enough such that the resulting hard gamma-ray spectra from these regions exhibit a normalisation that is large enough to make a substantial contribution to the total emission. Interestingly, as a consequence, the total gamma-ray spectrum conspires to mimic a steady-state configuration if integrated over the whole galaxy. Only considering the different radial contributions to the total emission spectra reveals the effect of spectrally resolved CR diffusion, which does not resemble the emission from the corresponding steady-state spectra in any radial bin if considered individually.

In the simulation with the most massive halo \texttt{M1e12-spec}, the hard gamma-ray spectra from the quickly outwards diffusing high-energy CRs dominate the total spectrum. This leads to a harder total gamma-ray spectrum than expected from steady-state. Furthermore, the gamma-ray spectra in the considered radial bins are in almost all regions harder than in a steady-state with the exception of the very central region ($R<2$~kpc).

To enable a more detailed comparison of the effect of spectrally resolved CR transport on the spectral shape of the resulting hadronic gamma-ray emission, we examine the emission spectral indices in the next section.

\section{\texorpdfstring{Comparison to observations and the interpretation of $\delta$}{Comparison to observations and the interpretation of delta}}
\label{sec: comparison to observations and interpretation of delta}

\subsection{Spectral index of gamma-ray emission} \label{sec: spectral index of gamma-ray emission}

\begin{figure*}
    \centering
    \includegraphics{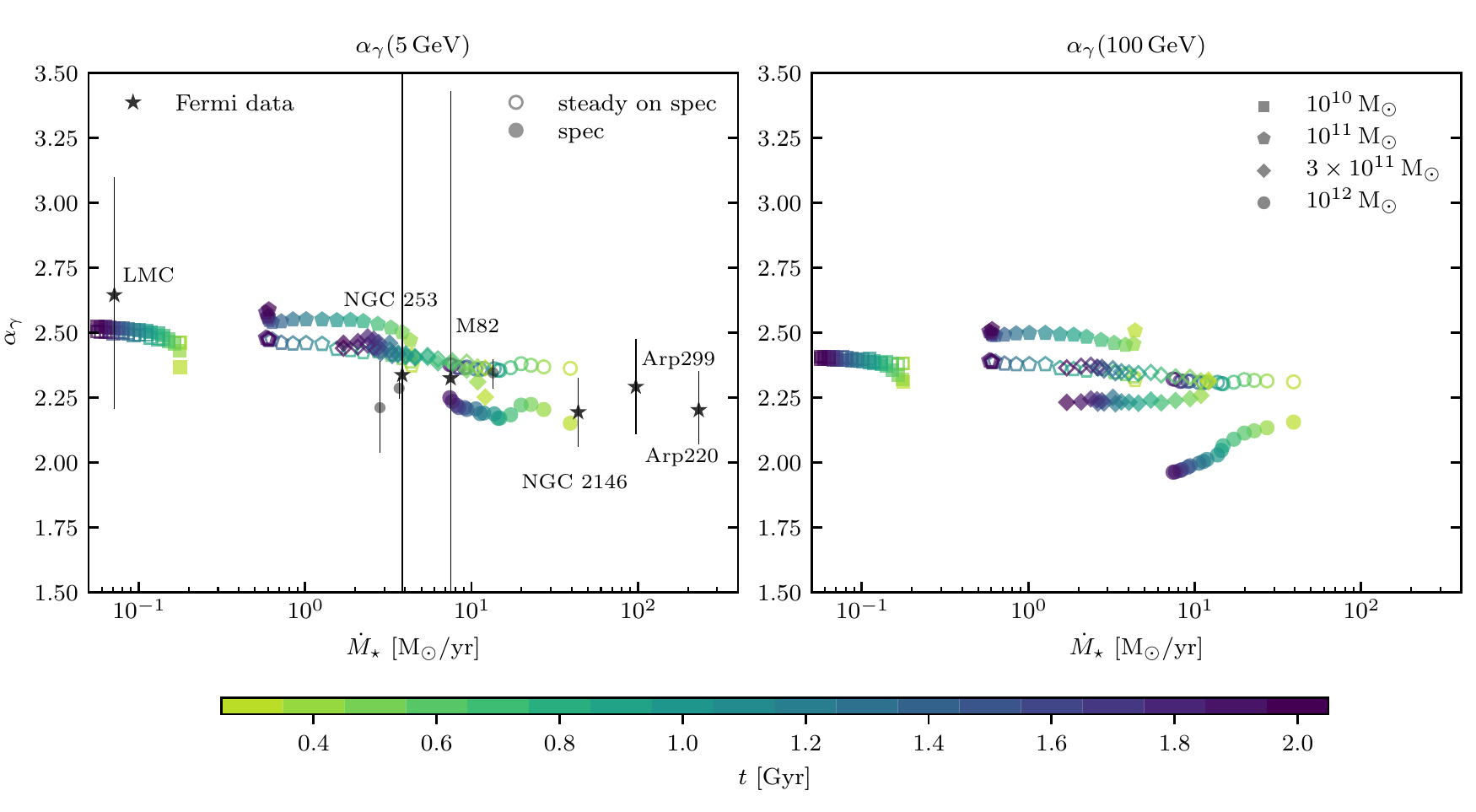}   
    \caption{We show the spectral indices of the hadronic gamma-ray spectra $\alpha_\gamma$ at 5~GeV (left-hand panel) and 100~GeV (right-hand panel) of our simulated galaxies, where the colour indicates the time of evolution. The different symbols correspond to different halo masses. 
    The left-hand panels additionally show the spectral indices of observed galaxies from \citet{2022Nunez-Castineyra}, where we differentiate between purely star-forming galaxies (star symbols) and galaxies with some potential AGN contribution to the gamma-ray emission (grey circles; see text for details). Filled symbols represent the spectral indices calculated from the spectrally resolved CR protons (model `spec'), while open symbols represent our steady-state model applied to the same simulations (model `steady on spec').}
    \label{fig: spectral-index - temporal evolution}
\end{figure*}

We calculate the spectral index $\alpha_\gamma$ of the hadronic gamma-ray spectra at 5 and 100~GeV, respectively, for all simulated halos at all times and show the result as a function of SFR in Fig.~\ref{fig: spectral-index - temporal evolution} (the time of the simulations are indicated by the same colors as in Fig.~\ref{fig:gamma-ray-spectra-time-evolution}). 
While we see a substantial evolution of the spectral indices at both energies shown for the spectrally resolved CR treatment (filled colored symbols), the steady-state model yields only a slight softening in the spectral index with time (i.e.\ decreasing SFR) for all halo masses.\footnote{Note that the steady-state model applied to the grey runs (model `grey') yields almost identical spectral indices in comparison to the steady-state spectra from the spectrally resolved runs (model `steady on spec').} 
For $\alpha_\gamma$ at 5~GeV, we observe that the spectra generally harden with increasing SFR in the spectral CR simulations. On the other hand, there is no such trend in $\alpha_\gamma$ at 100~GeV. Indeed, the two most massive halos (\texttt{M3e11-spec} and in particular \texttt{M1e12-spec}) exhibit a softening of their spectral index with increasing SFR, contrary to the general trend pointed out for 5~GeV. This is also clearly visible in the temporal evolution of the gamma-ray spectra shown in Fig.~\ref{fig:gamma-ray-spectra-time-evolution}, where the star-formation rate decreases with time after their initial peak.

As a caveat, we note that while at 5~GeV the hadronic gamma-ray emission is expected to dominate over leptonic contribution from inverse Compton (IC) or bremsstrahlung emission, at 100~GeV we expect a potential contribution from IC emission to the spectrum, which we do not model here. Including this might modify the spectral shapes and dilute the strong variations in the spectral indices that we obtain from neutral pion decay alone. To account for this properly \citep[beyond a steady-state approach as performed in ][]{2021WerhahnII} this requires a careful modelling of the time evolution of CR electron spectra, which goes beyond the scope of this paper and is left to future work.

In addition to our simulations, we show in Fig.~\ref{fig: spectral-index - temporal evolution} the spectral indices of a sample of nearby galaxies at 5~GeV from \citet{2022Nunez-Castineyra}, who extracted the spectral indices obtained from the most recent gamma-ray data from Fermi LAT, i.e.\ the DR3 version of the 4FGL catalogue \citep{2022FermiLAT}.

While the gamma-ray emission from the LMC, NGC~253, M82, NGC~2146, Arp229 and Arp220 is probably dominated by emission resulting from star-formation activity, we note that some galaxies of the sample are suspected to exhibit AGN activity. As pointed out in \citet[][and references therein]{2020Ajello}, the gamma-ray emission from NGC~3424 exceeds the calorimetric limit which could potentially be due to AGN contributions to the gamma-ray emission. Similarly, the gamma-ray emission from NGC~4945 and NGC~1069 might partly arise from the central AGN \citep[as discussed in][]{2020Ajello}. Consequently, we mark those galaxies with different symbols (grey circles) in Fig.~\ref{fig: spectral-index - temporal evolution}. 

Overall, we find that the differences in the modelling of the CR proton spectra are still smaller than the constraints provided by observations, where the error bars of the sample of star-forming galaxies extend well beyond the differences arising from our various modelling of the CR spectra, i.e. the spectral treatment vs. a steady-state configuration. Only the observed $\alpha_\gamma$ of NGC~2146 slightly favours the 
`spec' model that exhibits a harder spectrum in comparison to the steady-state approach.
We caution, however, that we expect this result to strongly depend on the model choice of $\alpha$ and $\delta$.

\begin{figure*}
    \centering
        \includegraphics{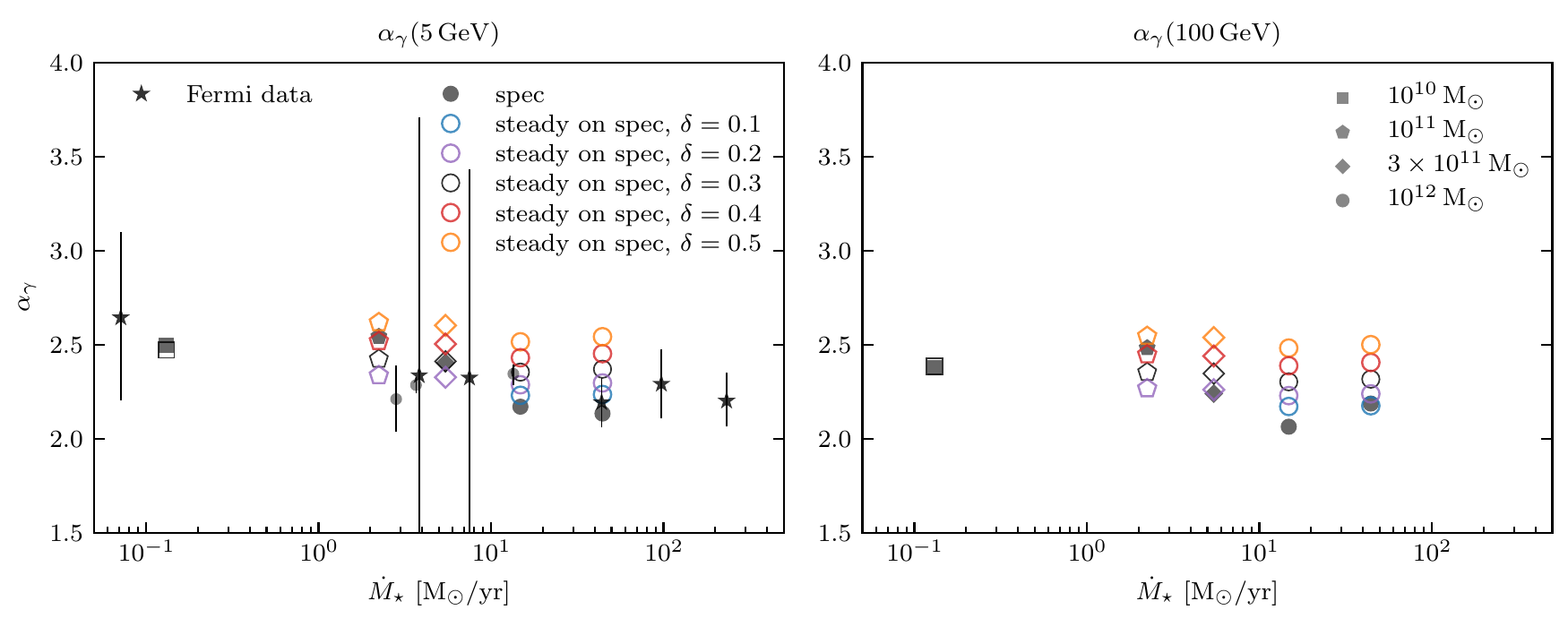}   
    \caption{Same as Fig.~\ref{fig: spectral-index - temporal evolution}, but only for a subset of the snapshots of our simulations for simplicity, i.e.\ snapshots at $t=0.8$~Gyr and in addition a snapshot of the \texttt{M1e12-spec} simulation with a SFR close to NGC~2146. We show the effect of varying the energy dependence of the diffusion coefficient in the steady-state model (`steady on spec'), which we vary from $\delta=0.1$ to $\delta=0.5$ (see different colors in the legend) on top of the simulations that adopt explicit energy dependent diffusion with $\delta=0.3$ (model `spec').}
    \label{fig: spectral-index - different deltas}
\end{figure*}

The deviations in the shapes of the gamma-ray spectra from spectrally resolved CR transport in comparison to a steady-state model suggest that it would be prudent to analyse the energy dependence of the diffusion coefficient in our steady-state model. 
Thus, in Fig.~\ref{fig: spectral-index - different deltas} we vary $\delta$ in the `steady on spec' model from 0.1 to 0.5 (the spectrally resolved CR runs were performed with $\delta=0.3$). 
We find that the simulated dwarf galaxy \texttt{M1e10-spec} matches a steady-state configuration with the same energy dependence ($\delta=0.3$) in terms of the spectral shape of the gamma-ray emission both at 5 and 100~GeV, respectively. 
Similarly, the \texttt{M3e11-spec} simulation matches the steady-state model with $\delta=0.3$. However, as discussed in Section~\ref{sec: Radial contribution to gamma-ray spectra}, this is only the case if we consider the total emission, while there is no region that would individually resemble the steady-state configuration with the same adopted $\delta$.

In contrast, Fig.~\ref{fig: spectral-index - different deltas} clearly shows that the spectral index of gamma-ray emission from the \texttt{M1e11-spec} simulation is more closely represented by a steady-state with a stronger energy dependence, close to $\delta=0.4$. In the case of the most massive halo, \texttt{M1e12-spec}, a shallower energy dependence for diffusion would be needed in order to reproduce the spectral shape of the more sophisticated spectrally resolved CR transport scheme.

The differences are identical for the 5 and 100 GeV spectral indices in the `steady on spec' model, while the `spec' model differs slightly in the case of each energy. In particular, the `spec' model of the \texttt{M3e11-spec} simulation exhibits a flatter spectral index in gamma-ray emission at 100~GeV in comparison to 5~GeV and resembles a steady-state with $\delta=0.2$.

\subsection{Gamma-ray spectra of starburst galaxies}
\label{sec: Gamma-ray spectra of starburst galaxies}

\begin{table*}
 \caption{Summary of the starburst galaxies that we compare our simulated spectra to in Figs.~\ref{fig: gamma-ray spectra NGC2146,M82} and \ref{fig: gamma-ray spectra NGC253}. The SFRs by \citet{2020Kornecki} have been calculated using far UV \citep{2007GilDePaz,2012Cortese} and IRAS $25\,\mathrm{\umu m}$ data \citep{2003Sanders}, whereas \citet{2022Nunez-Castineyra} combine the fluxes recorded in the four IRAS bands taken from the \citet{2003Sanders} and \citet{1990Moshir} source catalogues. In the column $L_\gamma$(\texttt{spec}), we present the gamma-ray luminosities integrated from 0.1 to 100~GeV obtained from the `spec' model as well as from the `steady on spec' model.}
 \label{Table-Galaxies-New}
 \begin{threeparttable}[t]
 \begin{tabular}{lcccccccc}
  \hline
  Galaxy & SFR (obs) &SFR (\texttt{spec}) &SFR (\texttt{grey}) & $L_{\gamma}$ (obs)& $L_\gamma$ (\texttt{spec}) & $L_\gamma$ (\texttt{grey}) & spectral simulation & grey simulation\\
  & $[\mathrm{M}_{\odot}~\mathrm{yr}^{-1}]$& $[\mathrm{M}_{\odot}~\mathrm{yr}^{-1}]$ &$[\mathrm{M}_{\odot}~\mathrm{yr}^{-1}]$ & $[10^{39}\mathrm{erg~s}^{-1}]$ & $[10^{39}\mathrm{erg~s}^{-1}]$ &$[10^{39}\mathrm{erg~s}^{-1}]$ & \texttt{name}, $t\,[\mathrm{Gyr}]$ & \texttt{name}, $t\,[\mathrm{Gyr}]$  \\
  \hline
    \hline
    NGC~2146 & $14.0\pm 0.5$\tnote{1} & $13.89$ & $14.00$ & $\left(88.6\right)$\tnote{3} & 46.35 / 59.20 & 45.12 & \texttt{M1e12-spec}, $0.98$ 
    & \texttt{M1e12-grey}, $1.15$ \\ 
            & $44.2\pm 12.6$\tnote{2}& $44.43$ & $44.06$ & $\left(86.2\pm 24.7\right)$\tnote{2} & 198.0 / 259.2 &  148.7
    & \texttt{M1e12-spec}, $0.27$ 
    & \texttt{M1e12-grey}, $0.30$\vspace{0.15cm} \\ 
    
   M82      & $10.4\pm 1.6$\tnote{1} & $10.40$ & $10.22$ &  $\left(18.5\right)$\tnote{3}& 28.72 /  40.81 &  32.64
   & \texttt{M1e12-spec}, $1.48$ 
   & \texttt{M1e12-grey}, $1.58$\\
            & & $10.45$ & $10.43$ & &  26.61 / 30.40 & 34.57
    & \texttt{M3e11-spec}, $0.44$ 
    & \texttt{M3e11-grey}, $0.45$\\
            &  $7.5 \pm 0.7$\tnote{2}& $7.51$& $7.41$ &  $\left(8.80\pm 0.88 \right)$\tnote{2} & 19.31 / 25.04 & 23.93 
    & \texttt{M1e12-spec}, $1.97$ 
    & \texttt{M1e12-grey}, $2.03$\\
            & & $7.51$ & $7.55$ & & 16.33 / 18.82 & 23.76
    & \texttt{M3e11-spec}, $0.61$ 
    & \texttt{M3e11-grey}, $0.61$\vspace{0.15cm} \\
    
   NGC~253  & $5.03 \pm 0.76$\tnote{1}& $5.07$  & $5.04$ &  $\left(11.6\right)$\tnote{3} & 9.902 / 11.33 & 15.48 & \texttt{M3e11-spec}, $0.84$ 
   & \texttt{M3e11-grey}, $0.83$ \\ 
            & $3.8 \pm 0.4$\tnote{2}  & $3.83$ & $3.82$ & $\left(7.14\pm 0.71 \right)$\tnote{2} & 7.005 / 7.948 & 11.49
    & \texttt{M3e11-spec}, $1.07$ 
    & \texttt{M3e11-grey}, $1.02$\\ 

  \hline
 \end{tabular}
 \begin{tablenotes}
      \item[1] \citet{2020Kornecki}.
      \item[2] \citet{2022Nunez-Castineyra}
      \item[3] \citet{2020Ajello}
   \end{tablenotes}
 \end{threeparttable}
\end{table*}

\begin{figure*}
    \centering
    \includegraphics{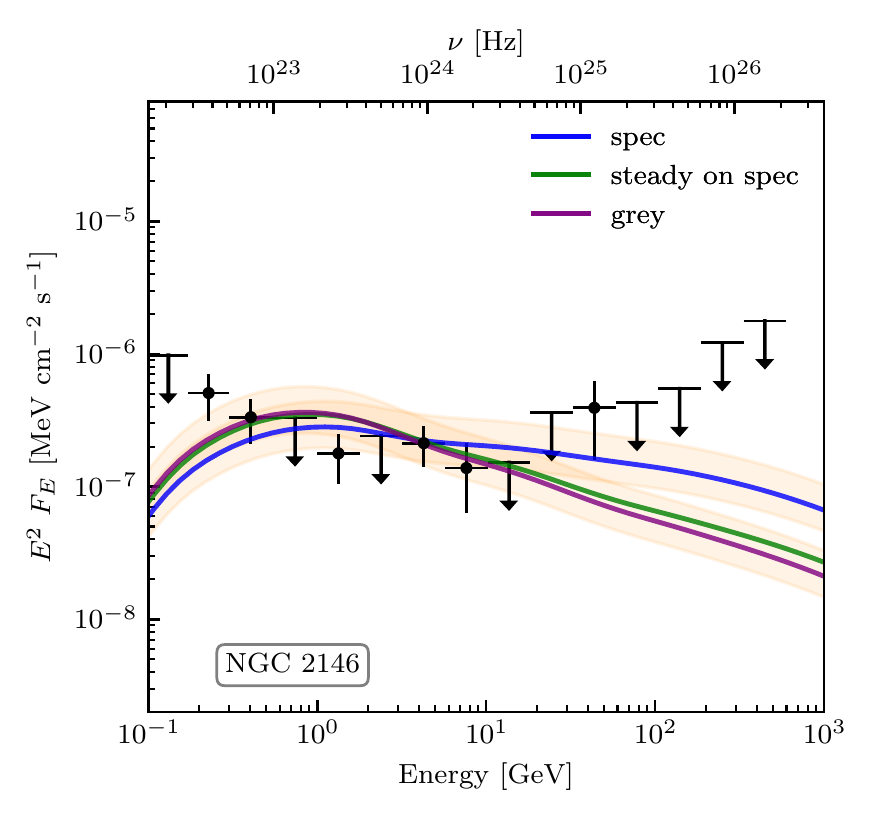}    \includegraphics{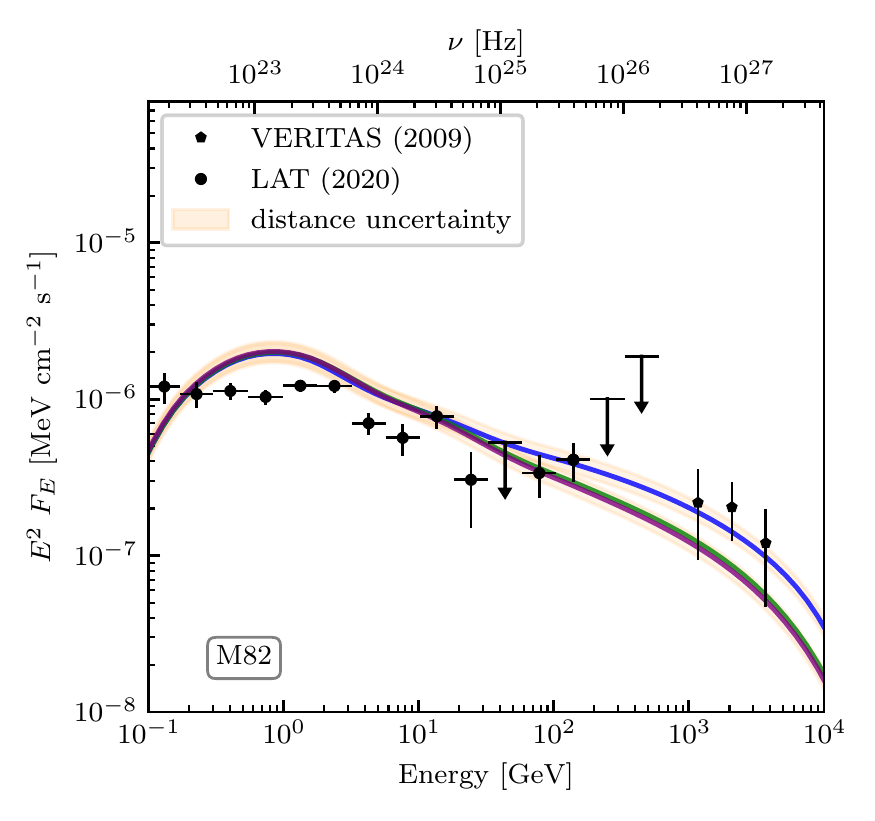}
    \caption{Observed gamma-ray spectra of NGC~2146 and M82 (black symbols, as indicated in the legend) together with the gamma-ray spectrum from neutral pion decay calculated from our simulations. For NGC~2146 (left-hand panel), we consider snapshots of the \texttt{M1e12-spec} and \texttt{M1e12-grey} simulations that are close to the observed SFR \citep[][see Table~\ref{Table-Galaxies-New}]{2022Nunez-Castineyra}. For the \texttt{M1e12-spec} simulation, we calculate the emission both from the spectrally resolved CR protons (`spec', blue line) and from our steady-state model applied to the same runs (`steady on spec', green line), while the resulting emission from the grey simulation is shown in purple. 
    We proceed in the same way for M82 (right-hand panel) but for the \texttt{M3e11-spec} and \texttt{M3e11-grey} simulations, respectively (see Table~\ref{Table-Galaxies-New}). We assume a distance of $d=18$~Mpc \citep{2012Adamo} for NGC~2146 (with a distance uncertainty of 20 per cent) and for M82 a distance of $d=3.53 \pm 0.25$~Mpc \citep{2020KourkchiCosmicflows4} for calculating the fluxes.
    To enable a visual comparison we re-scaled the simulated spectra to reproduce the observed gamma-ray luminosities (given in Table~\ref{Table-Galaxies-New}).}
    
    \label{fig: gamma-ray spectra NGC2146,M82}
\end{figure*}

\begin{figure*}
    \centering
    \includegraphics{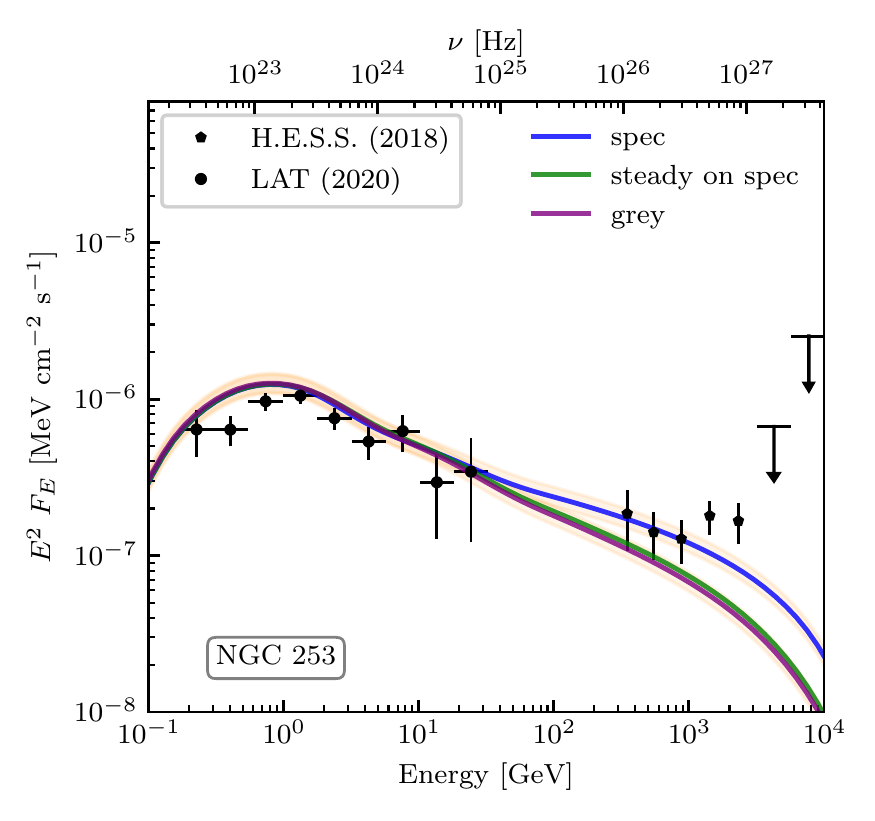}    \includegraphics{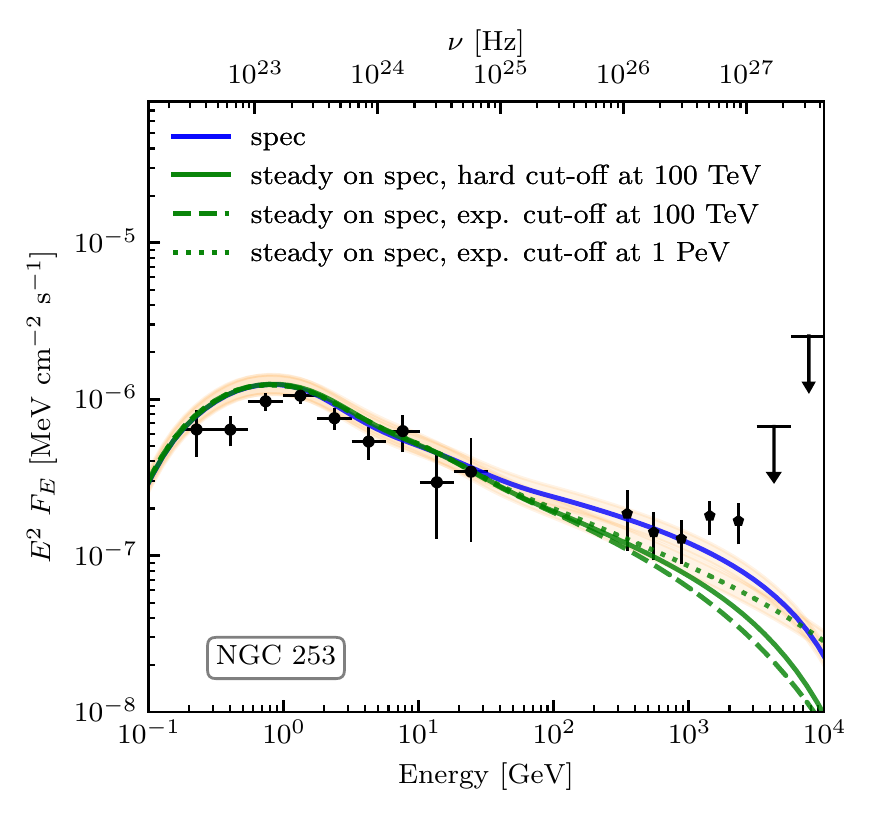}
    \caption{Same as Fig.~\ref{fig: gamma-ray spectra NGC2146,M82}, but for NGC~253 together with the simulated spectra from snapshots of the \texttt{M3e11-spec} and \texttt{M3e11-grey} simulations that are chosen such that they have a similar SFR to NGC~253 \citep[][see Table \ref{Table-Galaxies-New}]{2022Nunez-Castineyra}. We assume a distance of $d=3.56\pm 0.25$~Mpc \citep{2020KourkchiCosmicflows4} for calculating the fluxes. The right-hand panel shows the model `steady on spec' for different cut-offs in the injected CR proton spectrum. }
    \label{fig: gamma-ray spectra NGC253}
\end{figure*}

Next, we compare our simulated spectra to observed gamma-ray spectra of the star-forming galaxies NGC~2145, M82 and NGC~253 in Figs.~\ref{fig: gamma-ray spectra NGC2146,M82} and \ref{fig: gamma-ray spectra NGC253}. For all three galaxies, we plot the observations by Fermi-LAT \citep{2020Ajello} and additionally include data from \citet{2018HESS_NGC253} for NGC~253 and from \citet{2009VERITAS_M82} for M82.
As summarized in Table~\ref{Table-Galaxies-New}, we chose snapshots from our simulations that lie close to the observed galaxies in terms of their SFRs, where we both consider the values from \citet{2020Kornecki} and \citet{2022Nunez-Castineyra}, respectively. We analyse the spectrally resolved CR and the grey simulations. For the former, we additionally calculate the emission from the `steady on spec' model. In all considered cases, we obtain total gamma-ray luminosities (integrated from 0.1 to 100~GeV) that deviate at most a factor of three from the observed values (see Table~\ref{Table-Galaxies-New}).

We re-scale the simulated gamma-ray spectra to the observed total luminosities in order to enable a direct comparison of the spectral shapes with the data in Figs.~\ref{fig: gamma-ray spectra NGC2146,M82} and \ref{fig: gamma-ray spectra NGC253}. Here, we chose the snapshots from Table~\ref{Table-Galaxies-New} that resemble the SFRs cited in \citet{2022Nunez-Castineyra}.
For NGC~2146, we under-predict the observations in the sub-GeV regime. This could potentially be due to the lack of modelling of a contribution from non-thermal bremsstrahlung emission \citep[see][]{2021WerhahnII}. 
However, we obtain excellent agreement between the observed spectrum of NGC~253 and our spectrally resolved CR simulation (left-hand panel of Fig.~\ref{fig: gamma-ray spectra NGC253}), while the steady-state model (applied both to the spectral and grey runs) is softer towards the TeV regime.
Similarly, the spectrum of M82 is almost perfectly reproduced by our spectrally resolved CR proton simulation \texttt{M3e11-spec} (right-hand panel of Fig.~\ref{fig: gamma-ray spectra NGC2146,M82}), where again the steady-state spectra are softer at high gamma-ray energies. 
However, we find that the observed spectrum of M82 is only matched with our \texttt{M3e11-spec} simulation but not with a corresponding snapshot of the more massive halo \texttt{M1e12-spec} where the simulated galaxy exhibits a hard spectrum above $\sim10$~GeV in contradiction to the data.
A different estimation for the stellar mass of M82 obtained via the relation given by \citet{2013Cappellari} (which assumes that the dynamical mass within the observed region is $\approx M_\star$) using the K-band apparent magnitude $m_K = 4.665$ \citep{2006Skrutskie} and a distance of 3.5~Mpc (3.7~Mpc) yields a stellar mass of $M_\star \approx 4.1\times 10^{10}\,\msun$ ($4.5\times 10^{10}\,\msun$) and a halo mass of $\approx 1.4 \times 10^{12}\,\msun$ ($ 1.6 \times 10^{12}\,\msun$), where we used the relation by \citet{2010Moster} to infer the latter. 
However, \citet{2012Greco} estimate the dynamical mass of M82 (out to a radius of 4~kpc) to be only $\approx 10^{10}\,\msun$. If one assumes this to be an estimate for the stellar mass, this would imply a much smaller halo mass of $\approx 4.5\times 10^{11}\,\msun$.

In this section, we used for the source function in the `steady on spec' and the `grey' models a hard-cutoff in momentum space at 100~TeV to be consistent with the spectrally resolved method. 
Therefore, in the right-hand panel of Fig.~\ref{fig: gamma-ray spectra NGC253} we examine the effect of different cut-off momenta and shapes of the injected spectrum of CR protons on the resulting gamma-ray emission spectrum. This is important because the novel spectrally resolved scheme is computationally expensive and hence only allows for a limited number of momentum bins that are here chosen to range from 100~MeV~$c^{-1}$ to 100~TeV~$c^{-1}$. 
To assess the robustness of the artificially-introduced hard cut-off in momentum space, we consider two different functional forms for the source function for CR protons in our steady-state approach. In addition to the exponential cut-off with $q(p)\propto p^{-\alpha_\rmn{inj}} \exp(-p/p_\rmn{cut})$ (see Eq.~\ref{eq: source function steady-state q(p)}), we now also calculate the spectra by assuming a hard cut-off where $q(p)\propto p^{-\alpha_\rmn{inj}} \theta(p_\mathrm{cut}-p)$, with the Heaviside step function $\theta(p)$.

Clearly, the shape of the cut-off (either exponential or hard) only has a minor effect on the resulting gamma-ray spectrum if the same cut-off momentum $p_\rmn{cut}=100~\rmn{TeV}\,c^{-1}$ is assumed. 
However, the choice of a higher proton cut-off momentum at $p_\rmn{cut}=1~\rmn{PeV}\,c^{-1}$ leads to a harder gamma-ray spectrum that starts to deviate at gamma-ray energies $\gtrsim$~TeV from the spectrum obtained from setting $p_\rmn{cut}=100~\rmn{TeV}\,c^{-1}$.
Hence, the gamma-ray spectrum of the spectrally resolved model with a hard proton momentum cut-off at $100~\rmn{TeV}\,c^{-1}$ is robust for calculating gamma-ray emission up to $\sim$~TeV energies, whereas the artificial cut-off in momentum space softens the spectrum at higher gamma-ray energies. We anticipate that including further momentum bins above $100~\rmn{TeV}\,c^{-1}$ in the proton spectra of our spectrally resolved simulations would potentially harden the resulting gamma-ray spectra slightly above TeV energies.

The gamma-ray spectrum of extreme starbursts like Arp~220 has been suggested to soften above TeV energies because of pair production from the interaction of gamma-rays with infrared photons from the intense radiation field in the nucleus \citep{2004Torres, 2015Yoast-Hull, 2019Peretti}. Therefore, for a detailed modelling of very high-energy emission above a few TeV energies within such extreme environments, this effect should additionally be taken into account.


\subsection{The FIR-gamma-ray relation}

\begin{figure*}
\includegraphics[]{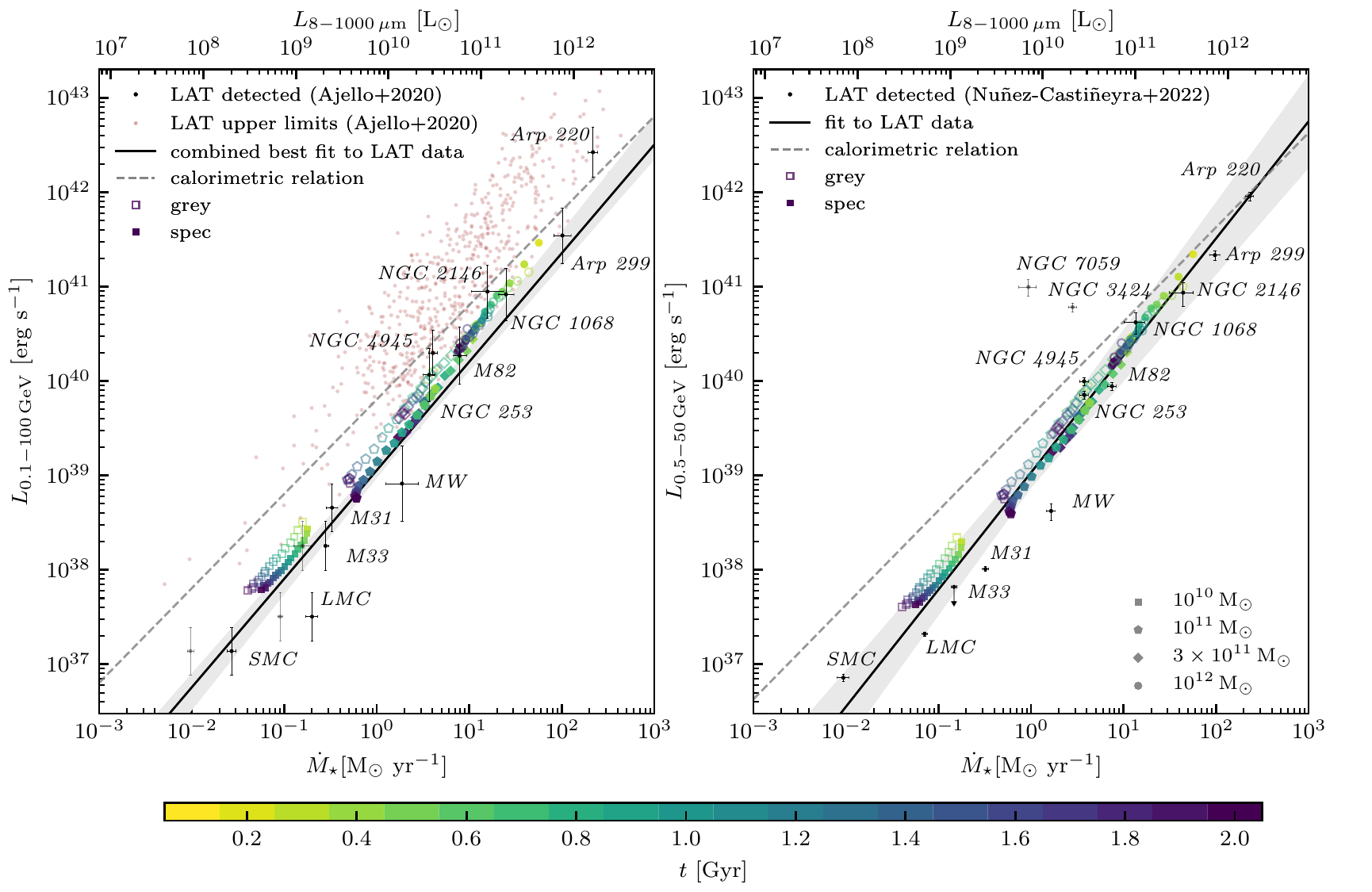}
\caption{Relation between the SFR and gamma-ray luminosity in different energy bands. The left-hand panel shows $L_{0.1-100\,\mathrm{GeV}}$ from neutral pion decay of our simulated galaxies (coloured symbols) together with the LAT data from \citet{2020Ajello}, whereas the right-hand panel shows the luminosities for a narrower energy band $L_{0.5-50\,\mathrm{GeV}}$ in order to compare it to more recent LAT observations from \citet{2022Nunez-Castineyra}. We show the luminosities from the `grey' (open coloured symbols) and `spec' model (filled coloured symbols) for all simulated halo masses (see the corresponding symbols in the legend) that are colour-coded by the time of the snapshot, starting from the peak of SFR in time intervals of 0.1~Gyr, respectively. From the snapshot with the highest SFR, we normalise the calorimetric relation (dashed line) via
$L_{E_1-E_2}/\eta_\rmn{cal,p}$ for $[E_1,E_2]=[0.1,100]$~GeV and $[0.5,50]$~GeV, respectively, where $\eta_\rmn{cal,p}$ is calculated from equations~(9) and (10) of \citet{2021WerhahnII}.
}
\label{fig: FIR-gamma-ray relation}
\end{figure*}

Finally, we compare the gamma-ray luminosities from neutral pion decay for all our simulated galaxies, which we obtain from Eq.~\ref{eq: L_(E1-E2)}, to observed data integrated over different energy bands in Fig.~\ref{fig: FIR-gamma-ray relation}.

\subsubsection{Observational data}
\citet{2020Ajello} analysed 10 years of Fermi LAT data in the energy band $0.1-100$~GeV, while more recent work by \citet{2022Nunez-Castineyra} extracted gamma-ray luminosities in the $0.5-50$~GeV band from 12 years of observations with Fermi LAT. 
The energy range of the latter is chosen to reduce potential leptonic contributions to the gamma-ray emission, i.e.\ from bremsstrahlung and IC emission from CR electrons at low and high gamma-ray energies, respectively. 
Besides the different energy integration ranges and data collection time, both data sets use different distances in some cases and the inferred FIR-luminosities and SFRs deviate for some of the observed galaxies. Therefore, we compare our simulations to both approaches separately in Fig.~\ref{fig: FIR-gamma-ray relation}.
Because \citet{2020Ajello} only analyse the relation between the FIR and gamma-ray luminosities of their sample of star-forming galaxies, while we plot the gamma-ray luminosities against the SFRs of our simulated galaxies, we convert the observed FIR luminosities to SFRs via the \citet{1998Kennicutt} relation. For low-SFR galaxies like the small and large Magellanic clouds (SMC and LMC) and M33, this relation is expected to break down \citep[as discussed in ][]{2020Kornecki}. Therefore, we additionally show their SFRs as obtained separately by \citet{Thirlwall2020} for M33 ($0.28^{+0.02}_{-0.01}\,\msun \mathrm{yr}^{-1}$) and by \citet{2020Kornecki} for the LMC ($0.20\pm 0.03\,\msun \mathrm{yr}^{-1}$) and the SMC ($0.027\pm 0.003\,\msun\mathrm{yr}^{-1}$) in the left-hand panel of Fig.~\ref{fig: FIR-gamma-ray relation}.
\citet{2022Nunez-Castineyra} provide a fit to the SFR-gamma-ray relation (right-hand panel of Fig.~\ref{fig: FIR-gamma-ray relation}) where they derive the SFRs of their sample also via the \citet{1998Kennicutt} relation.
Therefore, the same uncertainties hold here for the SMC, LMC and M33.
\citet{2022Nunez-Castineyra} derive their fit to the data from all observed galaxies but excluding NGC~7059 whose association with the observed LAT source has been doubted. We note that \citet{2020Ajello} additionally exclude NGC~3424 from their bonafide sample because it has been claimed to host an AGN \citep{2011Gavazzi} and shows some evidence of variability \citep{2019Peng}.

\subsubsection{Simulated gamma-ray to SFR relation}\label{sec: simulated gamma-ray to SFR relation}
In addition to the observational data, we present in Fig.~\ref{fig: FIR-gamma-ray relation} our simulated gamma-ray luminosities versus the SFRs of our simulated galaxies. We show all snapshots with a time difference of $\Delta t=0.1$~Gyr for all halo masses of both the spectral and grey simulations (filled and open symbols, respectively).
While we slightly overestimate the gamma-ray luminosities $L_{0.1-100\,\mathrm{GeV}}$ in comparison to the LAT data from \citet{2020Ajello}, we obtain an excellent agreement with the observed luminosities $L_{0.5-50\,\mathrm{GeV}}$ by \citet{2022Nunez-Castineyra}.
In both cases, we find that the `spec' model exhibits almost identical luminosities compared to the `grey' model at high SFRs, while it deviates towards lower gamma-ray luminosities with decreasing SFRs and halo masses. However, the differences are small in comparison to the observed scatter in the relation. 
We note that we adapt an acceleration efficiency of CRs at SNRs of $\zeta_\mathrm{SN}=0.10$, while \citet{2021WerhahnII} previously found that an injection efficiency of $\zeta_\mathrm{SN}=0.05$ in their grey simulations better reproduces the observed relation from \citet{2020Ajello}. In the previous approach, however, the authors model in addition to CR protons the steady-state spectra of CR electrons and calculate their resulting bremsstrahlung and IC emission. 
The contribution of neutral pion decay to the total gamma-ray luminosity $L_{0.1-100\,\mathrm{GeV}}$ in \citet{2021WerhahnII} is around 60 per cent at $\dot{M}_\star \gtrsim1~\msun \mathrm{yr}^{-1}$ and decreases to $\sim$40 per cent at $\dot{M}_\star \sim 10^{-2}~\msun \mathrm{yr}^{-1}$. The choice of a narrower energy band of 0.5-50~GeV is expected to reduce the leptonic contribution further. 
Nevertheless, we conclude that our hadronic gamma-ray luminosities in both the `grey' and `spec' models are probably uncertain by roughly a factor of two due to the choice of $\zeta_\mathrm{SN}=0.10$ and the lack of accounting for leptonic gamma-ray emission in this work. Both uncertainties therefore have an opposite effect on the gamma-ray luminosities and hence may well approximately cancel each other out.

Additionally, we analyse the calorimetric relation in comparison to the observed and simulated FIR-gamma-ray relations in Fig.~\ref{fig: FIR-gamma-ray relation}. This relation is supposed to represent the limit where CR protons would lose all of their energy due to neutral pion decay. In order to estimate the normalisation of the calorimetric relation from our simulations, we calculate the calorimetric fraction $\eta_\rmn{cal,p}$ from the snapshot with the highest SFR according to equations~(9) and (10) of \citet{2021WerhahnII}. This definition takes into account that only a fraction of the CR proton population is able to produce gamma-rays in the considered energy range, denoted as the bolometric energy fraction $\zeta_\rmn{bol}$. 
While in \citet{2021WerhahnII}, it was found that for the range $0.1-100$~GeV $\zeta_\rmn{bol}\approx 0.6$, we find $\zeta_\rmn{bol}\approx 0.4$ in the energy range $0.5-50$~GeV. This results in a calorimetric fraction of our snapshot with the highest SFR and gamma-ray luminosity in the `spec' model of $\eta_\rmn{cal,p}=0.81$ ($\eta_\rmn{cal,p}=0.92$) for the energy band $0.1-100$~GeV ($0.5-50$~GeV), suggesting that the energy range $0.1-100$~GeV of emitted gamma-ray photons is less calorimetric than the narrower band. 
This is because the broader band up to 100~GeV traces higher energetic CR protons which are more affected by diffusive losses due to the inclusion of energy dependent diffusion.

For both relations, we find that they are consistent with the observed data; highly SF galaxies are close to the calorimetric relation, whereas the luminosities deviate from the relation for galaxies with decreasing SFRs. This has also been found in previous work \citep{2007Thompson,2011Lacki,2012AckermannGamma, 2014Martin, 2017bPfrommer, 2020Kornecki, 2021WerhahnII}.

\section{Discussion and caveats}\label{sec: discussion and caveats}

Whilst we believe that our conclusions are robust, there are, naturally, some associated caveats to our results. For example, whilst the simulated gamma-ray spectra that we compare to star-forming galaxies NGC~2146 and M82 in Sec.~\ref{sec: Gamma-ray spectra of starburst galaxies} are broadly consistent with a number of previous works that model steady-state spectra within one-zone models \citep[e.g.\ ][]{2010Lacki, 2013Yoast-Hull, 2019Peretti}, we note that we do not include leptonic contributions to the gamma-ray emission in this work. Including non-thermal bremsstrahlung and IC emission might flatten the gamma-ray spectra below the pion decay bump, although this contribution has been found to be subdominant for the total emission of star-forming galaxies \citep[e.g.\ ][]{2005Domingo-SantamariaTorres, 2007Thompson, 2008Persic,2009deCeaDelPozo, 2012Paglione,2013Yoast-Hull} and in the Milky Way up to energies of $\sim10$~GeV \citep{2010Strong}. 
Furthermore, in line with the results in \citet{2021WerhahnII}, we expect the leptonic contribution to the gamma-ray luminosities in the energy band of 0.1 to 100~GeV for highly star-forming galaxies like M82, NGC~253 and NGC~2146 to be below 50 per cent and potentially even smaller in the narrower energy band of 0.5 to 50~GeV.
In contrast, IC emission might become more relevant at high energies $\gtrsim 10~$GeV in galaxies with smaller SFRs like the SMC \citep{2021WerhahnII}. 
Analysis of this parameter space is unfortunately outside the scope of this paper as an accurate modelling of the CR electron population beyond steady-state modelling is needed to account for a detailed calculation of bremsstrahlung and IC emission, the latter being particularly sensitive to the high-energy shape of the CR electron spectrum. We therefore postpone the modelling of live CR electron spectra to future work with the \crest code \citep[][]{2019Winner}.
Coupling this to the non-thermal emission code \crayon described in \citet{2021WerhahnIII,2021WerhahnI,2021WerhahnII} will enable a detailed calculation of the leptonic gamma-ray emission of star-forming galaxies and enable a more robust conclusion about the role of leptonic emission in the gamma-ray regime beyond the steady-state approach.

Similarly, our simulated FIR-gamma-ray relation is in agreement with the observed relation by \citet{2020Ajello}, as well as the data and modelling by \citet{2022Nunez-Castineyra} (see Fig.~\ref{fig: FIR-gamma-ray relation}), with the former also matched in previous studies of isolated galaxies \citep{2017bPfrommer,2021WerhahnII} and within cosmological settings \citep{2020Buck}. As discussed in Section~\ref{sec: simulated gamma-ray to SFR relation}, however, \citet{2021WerhahnII} additionally account for IC and bremsstrahlung emission from CR electrons and subsequently find a lower acceleration efficiency of $\zeta_\mathrm{SN}=0.05$ is required to match the observed relation.

Meanwhile, the FIRE-2 simulations by \citet{2019Chan} require a diffusion coefficient of $\gtrsim 3\times 10^{29}\,\mathrm{cm^2\,s^{-1}}$ in order to match the observed relation with their simulations. This is, however, in tension with the results of the aforementioned simulations \citep{2017bPfrommer, 2020Buck, 2021WerhahnII, 2022Nunez-Castineyra} and the work presented in this paper, which all match the observed gamma-ray data with smaller diffusion coefficients.

Another uncertainty relating to this work is the fixed value of the energy dependence of the diffusion coefficient $\delta$, which we chose to be 0.3 in all spectrally resolved simulations analysed here. 
Varying this value may be particularly important for a proper comparison with observations from the Milky Way, where the detected ratios of beryllium isotopes suggest $\delta=0.5$ \citep{2020bEvoli}. We aim to explore a variation of $\delta$ in the `spec' model in future work. This will additionally require running the simulation with a Milky Way-like halo mass to much later times, when the SFR has sufficiently decreased to the observed value of $1.7~\msun~\mathrm{yr}^{-1}$ \citep{2011Chomiuk}. 

Finally, we note that in our equations CR diffusion is modelled by means of a spatially constant diffusion coefficient, rather than being self-consistently calculated in the picture of CR self-confinement or extrinsic confinement in MHD turbulence.
It has been shown previously that the excitation of turbulence by CRs escaping a shock upstream has a non-linear effect on the escape and confinement of CRs in the region around their sources \citep[see e.g.,][and references therein]{2004Bell, 2013Bell, 2021Marcowith}. This self-excited magnetic turbulence can reduce the diffusion coefficient, particularly in the vicinity of CR sources \citep{2008Reville, 2021Schroer, 2022Recchia}. This depends, however, on the properties of the ISM surrounding the source \citep{2007Reville, 2021Reville, 2016Nava, 2019Nava}. Additionally, a reduced diffusion coefficient around the sources can also impact the phase-space structure of star-forming regions so that the resulting larger CR pressure prevents local gas fragmentation even in cases when the disc is unstable, thereby helping to maintain a regular grand-design spiral structure \citep{2021Semenov}.
With this in mind, a further improvement in our modelling could be achieved by adapting a refined model of CR transport in the two-moment approach that accounts for a temporarily and spatially varying diffusion coefficient in the self-confinement picture \citep{2018Jiang,2019ThomasPfrommer, 2022ThomasPfrommer, 2021Thomas}. 
This could potentially transform the morphology of our high-energy emission maps at $\sim100$~GeV to become more source dominated. 




\section{Conclusions}\label{sec: discussion and conclusion}

In this work, we simulate the observational signatures of gamma-ray emission that arise from neutral pion decay for galaxies with halo masses ranging from $10^{10}$ to $10^{12}\,\msun$. To this end, we analyse a series of MHD simulations of isolated galaxies from \citet{2023Girichidis} using the moving-mesh code \arepo, where we include a novel extension of the code that accounts for spectrally resolved CR transport \citep{2020Girichidis,2022Girichidis}. Instead of only following the evolution of the momentum-integrated energy density, as employed in the grey approximation, the spectrally resolved method represents the CR distribution function as a piece-wise power law in momentum space for every computational cell, as well as allowing for energy-dependent diffusion.

To isolate the observational signatures that arise from this novel treatment, we compare the hadronic gamma-ray emission predicted by the spectrally resolved simulations (model `spec’) with that predicted by a steady-state model applied to the same simulations (model `steady on spec’) and with simulations run using the grey approximation (model `grey’). For the latter, only the total CR energy density is evolved within the simulations. Ultimately, this leads to only minor variations in the spatially resolved spectra and gamma-ray emission resulting from the `grey' and `steady on spec' models (see Fig.~\ref{fig:maps-gamma-ray-spec}). Indeed, the models are almost identical when averaged over the whole disc (see Fig.~\ref{fig: CR protons all halos}).
We therefore focus on the differences between our `steady on spec' and `spec' models in our conclusions. We find that:

\begin{itemize}
\item Whilst the spatial distribution of the total gamma-ray luminosity is almost identical in the `steady on spec’ and `spec’ models, the distribution of the luminosity in specific energy bands differs significantly (see Figs.~\ref{fig:maps-gamma-ray-spec} and \ref{fig:gamma-ray-diff-contribution}). In particular, due to the explicit modelling of spectrally resolved CR transport, high-energy CRs diffuse faster into the outskirts of the disc and hence the radial distribution of the high-energy gamma-ray emission is much more extended in the `spec’ model compared to the steady-state approach.

\item For the central regions of the simulated galaxies, the `spec’ model yields steeper CR spectra for all halo masses in comparison to the corresponding `steady on spec’ model. However, in the outskirts of the galaxy, the outwards diffusing high-energy CRs lead to much flatter spectra (see Fig.~\ref{fig:CR-proton-spectra-radialbins}). Indeed, for our largest simulated galaxy (with a halo mass of $10^{12}\,\msun$), the CR spectra in the outskirts actually rise as a function of $P$ in the `spec' model above $\sim1$~GeV, whilst the spectral slope is negative for the steady-state model.
Despite this, due to the varying gas distribution in the different galaxies, the hard CR spectra at large radii only contribute to the gamma-ray emission significantly in the `spec’ model of the two most massive halos, i.e.\ the \texttt{M3e11-spec} and \texttt{M1e12-spec} simulations, respectively. In the former, the combination of steep central and flat outer spectra coincide to closely resemble the total gamma-ray spectra of the corresponding `steady on spec’ model, whereas in the most massive halo, the `spec’ model already yields much harder gamma-ray spectra after a short run-time (see Figs.~\ref{fig:gamma-ray-spectra-time-evolution} and \ref{fig: gamma-ray spectra radial bins}).

\item We find the spectral indices of gamma-ray emission $\alpha_\gamma$ at 5~GeV generated by the different models match the observed data from nearby star-forming galaxies to within observational errors (see Fig.~\ref{fig: spectral-index - temporal evolution}). This implies that observations are not yet constraining enough to discriminate between our different methods of modelling CR spectra. 

\item We find that the shape of the gamma-ray emission (i.e.\ the spectral index $\alpha_\gamma$) from the `spec’ model at a fixed energy can be represented by a steady-state configuration if one allows for a variation in the energy dependence of the diffusion coefficient $\delta$ (see Fig.~\ref{fig: spectral-index - different deltas}).
For example, the \texttt{M1e11-spec} simulation with $\delta=0.3$ can be reproduced globally by adapting a steady-state model with a stronger energy dependence of $\sim 0.4$, whereas the spectral shapes of the total gamma-ray emission of the \texttt{M1e12-spec} simulation requires $\delta<0.2$ in the `steady on spec' model to reproduce the results from the `spec' model. 

\item The total gamma-ray spectra of the star-forming galaxy NGC~2146 is largely consistent with the hadronic emission from our `spec' model. Moreover, the observed spectra of NGC~253 and M82 are in excellent agreement with our `spec’ model, while the `steady on spec' and `grey' models exhibit softer spectra at TeV energies than suggested by the observed data (see Fig.~\ref{fig: gamma-ray spectra NGC2146,M82} and \ref{fig: gamma-ray spectra NGC253}).

\item In the FIR-gamma-ray relation (Fig.~\ref{fig: FIR-gamma-ray relation}), we find that the `spec' model yields similar gamma-ray luminosities when compared to the `grey' model at high SFRs but lower values at SFRs $\lesssim 10\,\msun\mathrm{yr}^{-1}$. The relation obtained from the `spec' model coincides particularly well with the most recent observations by Fermi LAT of the gamma-ray luminosities $L_{0.5-50~\mathrm{GeV}}$ of nearby star-forming galaxies \citep{2022Nunez-Castineyra}. 

\end{itemize}

We conclude that the steady-state approach is an appropriate approximation for calculating the gamma-ray emission from star-forming galaxies if, and only if, such analysis is performed globally. For each halo mass and stage of temporal evolution there exists a mapping of the necessary parameters such that one can approximate the total emission spectra predicted by the spectrally resolved model for a specific energy band with a steady-state model. Additionally, this mapping only requires slight variation with energy.
However, the spatial distribution of the gamma-ray emission at various energies differs significantly when modelling the energy dependent diffusion of CRs explicitly. 
Hence, we conclude that accounting for spectrally resolved CR transport is, in particular, essential for an accurate prediction of the spatial distribution of high-energy hadronic gamma-ray emission of star-forming galaxies. This will be highly relevant for future observations with the next generation Cherenkov Telescope Array observatory.

\section*{Acknowledgements}
MW, PG, and CP acknowledge support by the European Research Council under
ERC-CoG grant CRAGSMAN-646955 and ERC-AdG grant PICOGAL-101019746. PG also acknowledges funding from the ERC Synergy Grant ECOGAL (grant 855130). JW acknowledges support by the German Science Foundation (DFG) under grant 444932369.

\section*{Data Availability}
The data underlying this article will be shared on reasonable request to the corresponding author.

\appendix

\bibliographystyle{mnras}
\bibliography{literatur}

\end{document}